# Particle-Number Projection and the Density Functional Theory

J. Dobaczewski,[1,2,3,4,5] M.V. Stoitsov,[1,2,3,6] W. Nazarewicz,[1,2,4] and P.-G. Reinhard[7,3]

[1] *Department of Physics & Astronomy, University of Tennessee, Knoxville, Tennessee 37996, USA*
[2] *Physics Division, Oak Ridge National Laboratory, P.O. Box 2008, Oak Ridge, Tennessee 37831, USA*
[3] *Joint Institute for Heavy-Ion Research, Oak Ridge, Tennessee 37831, USA*
[4] *Institute of Theoretical Physics, Warsaw University, ul. Hoża 69, 00-681 Warsaw, Poland*
[5] *Department of Physics, P.O. Box 35 (YFL), FI-40014 University of Jyväskylä, Finland*
[6] *Institute of Nuclear Research and Nuclear Energy, Bulgarian Academy of Sciences, Sofia-1784, Bulgaria*
[7] *Institut für Theoretische Physik, Universität Erlangen, Staudtstrasse 7, D-91058 Erlangen, Germany*

In the framework of the Density Functional Theory for superconductors, we study the restoration of the particle number symmetry by means of the projection technique. Conceptual problems are outlined and numerical difficulties are discussed. Both are related to the fact that neither the many-body Hamiltonian nor the wave function of the system appear explicitly in the Density Functional Theory. Similar obstacles are encountered in self-consistent theories utilizing density-dependent effective interactions.

## I. INTRODUCTION

Superconductivity plays a central role in describing low-temperature properties of correlated many-fermion systems. Within the mean-field theory, fermionic pairing is best treated in the Hartree-Fock-Bogoliubov (HFB) [1] or Bogoliubov-de Gennes (BdG) [2] formalism. In the presence of superconducting condensate, the standard product state ansatz for the nuclear wave function breaks the particle-number (PN) symmetry [1, 3]. In principle, the broken symmetry needs to be restored, especially if one looks at observables that strongly depend on PN. The many-body correlations associated with the symmetry-breaking are particularly important for small systems where the finite-size effects are appreciable, such as atomic nuclei or metallic grains, or in the limit of weak pairing where pairing correlations have dynamic character.

For complex superconducting system,s a theoretical tool of choice is the Density Functional Theory (DFT) [4, 5]. The theory is built on theorems showing the existence of energy functionals for many-body systems, which include, in principle, all many-body correlations. The generalization of the DFT to the case of fermionic pairing was formulated for electronic superconductors in Refs. [6, 7, 8]. The resulting HFB/BdG equations can be viewed as the generalized Kohn-Sham equations of the standard DFT.

In the nuclear case, the DFT is the only tractable theory that can be applied across the entire table of nuclides. Historically, the first nuclear energy density functionals appeared long ago [9, 10, 11] in the context of the Hartree-Fock (HF) method used with zero-range, density-dependent interactions such as the Skyrme force. The main ingredient of the nuclear DFT [12] is the energy density functional that depends on densities and currents representing distributions of nucleonic matter, spins, momentum, and kinetic energy, as well as their derivatives (gradient terms). To account for nuclear superfluidity, the functional is augmented by the pairing term (see Ref. [13] for a review). The challenges faced by the nuclear DFT are: (i) the existence of two kinds of fermions; (ii) the essential role of pairing; and (iii) the need for symmetry restoration in finite, self-bound systems. The two latter points are of particular importance in the context of this study. The features (i) and (iii) are specifically nuclear; with very few exceptions, they are not present in the electronic Coulomb problem.

It is important to recall that the realistic energy density functional does not have to be related to any given effective Hamiltonian, i.e., an effective interaction could be secondary to the functional. This strategy is used in all modern nuclear DFT applications. In the absence of a Hamiltonian (and wave function), the restoration of spontaneously broken symmetries in DFT poses a conceptional dilemma [14, 15, 16, 17] and a serious challenge that needs to be properly addressed. One important question related to DFT for self-bound systems concerns the functional itself: how do you construct it in terms of intrinsic (body-fixed) densities? While it is possible to formulate the Kohn-Sham procedure in language of intrinsic densities [18, 19], the pathway to practical applications is still not clear.

Sticking to DFT for superconductors and PN symmetry, several schemes can be adopted. One is to formulate the theory in language of the usual (particle) density only, without explicitly invoking the anomalous (pair) density that is at the heart of the PN symmetry violation [20, 21]. Another strategy is to incorporate the PN restoration procedure into the DFT framework. This can be done by employing the generalized Wick's theorem (see, e.g., [22, 23, 24]). Recently, full PN projection before varia-tion has been carried out for the first time within the Skyrme-DFT framework employing zero-range pairing [23, 25, 26]. It was demonstrated that the resulting projected DFT equations (similar to the PN-conserving HFB equations originally proposed in Refs. [27, 28]) can be obtained from the standard Skyrme-HFB equations in coordinate space by replacing the intrinsic densities and currents by their gauge-angle-dependent counterparts. Using the variation-after-projection method, one can properly describe transitions between normal and superconducting phases in finite systems, which are inherent in atomic nuclei.

As mentioned above, the restoration of broken symmetries in the framework of DFT causes a number of questions, mainly related to the density dependence of the underlying interaction and to different treatment of particle-hole and particle-particle channels [22, 25, 29]. For instance, it has been realized for some time [22, 25, 29, 30, 31, 32] that the PN projection applied within the DFT framework is plagued with difficulties related to vanishing overlaps between gauge-rotated intrinsic states. This concerns any functional that uses density-dependent terms and thus is not related to an average of a Hamiltonian. In particular, the most frequently used approaches based on the Skyrme, Gogny, or relativistic-mean-field functionals all fall into this category.

In this study we investigate the analytic structure of the projected DFT, focusing on origins of difficulties. In recent works [33, 34], a way to remedy some of the problems has been proposed. The PN-projected Skyrme-DFT formalism employed in our work has been outlined in Ref. [23], and we follow their notation. Our manuscript is organized as follows. The analytic structure of the projected HFB is discussed in Sec. II. The DFT extension of the formalism is described in III. Numerical examples are contained in IV. Finally, Sec. V contains conclusions of this work.

## II. PARTICLE-NUMBER-PROJECTED HFB

In the context of HFB theory [1], the particle-number-projected (PNP) state is given by the standard expression

$$|\Psi_N\rangle \equiv \hat{P}_N|\Phi\rangle = \frac{1}{2\pi}\int_0^{2\pi} \mathrm{d}\phi\, e^{i\phi(\hat{N}-N)}|\Phi\rangle, \tag{1}$$

where $\hat{N}$ is the PN operator, $\phi$ is the gauge angle, $\hat{P}_N$ is the projection operator for $N$ particles, and $|\Phi\rangle$ is the HFB wave function (generalized product state) which does not have well-defined particle number. This expression, after the integral is discretized, is most often used in practical calculations. However, it only constitutes a specific realization of a more general form [35] given by the contour integral,

$$|\Psi_N\rangle \equiv \hat{P}_N|\Phi\rangle = \frac{1}{2\pi i}\oint_C \mathrm{d}z\, z^{\hat{N}-N-1}|\Phi\rangle, \tag{2}$$

where C is an arbitrary closed contour encircling the origin $z = 0$ of the complex plane.

### A. Shifted HFB states

Let us introduce several useful notations that will be used later. First, we call the operator appearing under the integral (2) the shift operator,

$$\hat{z}(z) = z^{\hat{N}} = e^{(\eta+i\phi)\hat{N}}, \tag{3}$$

parametrized by means of a single complex number $z$, $\ln(z)=\eta + i\phi$. The shift operator $\hat{z}(z)$ is parametrized by the complex number $z$ and constitutes a non-unitary Bogoliubov transformation (in fact, a non-unitary single-particle basis transformation) of simple kind, i.e.,

$$\begin{aligned}\hat{z}a_n^+\hat{z}^{-1} &= za_n^+,\\ \hat{z}a_n\hat{z}^{-1} &= z^{-1}a_n\end{aligned} \tag{4}$$

or

$$\begin{aligned}\hat{z}^{-1}a_n^+\hat{z} &= z^{-1}a_n^+,\\ \hat{z}^{-1}a_n\hat{z} &= za_n.\end{aligned} \tag{5}$$

Obviously, for $z = 1$, the shift operator is equal to identity.

Second, we define the shifted HFB states as

$$|\Phi(z)\rangle = \hat{z}(z)|\Phi\rangle. \tag{6}$$

When the HFB state $|\Phi\rangle$ is expressed through the Thouless theorem [1] (we assume an even number of particles for simplicity),

$$|\Phi\rangle = \mathcal{N}\exp\left(\tfrac{1}{2}\sum_{mn} Z^*_{mn}a_m^+a_n^+\right)|0\rangle, \tag{7}$$

the shifted HFB states read

$$|\Phi(z)\rangle = \mathcal{N}\exp\left(\tfrac{1}{2}z^2\sum_{mn} Z^*_{mn}a_m^+a_n^+\right)|0\rangle, \tag{8}$$

where $\mathcal{N}$ is the normalization constant of the HFB state (7). Similarly, for the HFB state expressed in the canonical basis or for a BCS state,

$$|\Phi\rangle = \prod_{n>0}\left(u_n + v_n a_n^+a_{\bar{n}}^+\right)|0\rangle, \tag{9}$$

the shifted state reads

$$|\Phi(z)\rangle = \prod_{n>0}\left(u_n + z^2\, v_n a_n^+a_{\bar{n}}^+\right)|0\rangle, \tag{10}$$

where $u_n$ and $v_n$ are the real HFB occupation amplitudes in the canonical basis and the product $\prod_{n>0}$ involves only one state from each pair of canonical partners (see Ref. [1] for details).

We call $\hat{z}(z)$ a shift, because it moves the HFB state $|\Phi\rangle$=$|\Phi(1)\rangle$ from its original position at $z = 1$ to a different point $z$ in the complex plane. Since consecutive shift transformations correspond to products of the shift parameters $z$, the parameters $\eta$ and $\phi$ in Eq. (3) are additive.

### B. Projected HFB states

The Thouless theorem (7) allows us to express the HFB state $|\Phi\rangle$ and shifted HFB state $|\Phi(z)\rangle$ as sums of components having different particle numbers,

$$|\Phi\rangle = \mathcal{N}\sum_{k=0}^{\infty}\frac{(\hat{Z}^{+})^{k}}{k!}|0\rangle, \tag{11}$$

$$|\Phi(z)\rangle = \mathcal{N}\sum_{k=0}^{\infty}z^{2k}\frac{(\hat{Z}^{+})^{k}}{k!}|0\rangle, \tag{12}$$

where $\hat{Z}^{+} = \frac{1}{2}\sum_{mn}Z^{*}_{mn}a^{+}_{m}a^{+}_{n}$ is the Thouless pair-creation operator. It then trivially follows that the shift transformation does not change any of the particle-number-projected states (2),

$$|\Psi_N\rangle = \mathcal{N}\frac{(\hat{Z}^{+})^{N/2}}{(N/2)!}|0\rangle, \tag{13}$$

but only scales the coefficients in the sum of Eq. (12).

Since the shifted states (12) are manifestly analytical in $z$, all closed contours $C$ in Eq. (2) give, by the Cauchy theorem, the same result. Among them, the integral in Eq. (1) simply corresponds to the unit circle.

The analyticity of $|\Phi(z)\rangle$ results in a simple and elegant representation of the projected state:

$$|\Psi_N\rangle \equiv \hat{P}_N|\Phi\rangle = \operatorname*{Res}_{z\,=\,0}\, z^{-N-1}|\Phi(z)\rangle. \tag{14}$$

Indeed, in the sum of Eq. (12), only the term with $N = 2k$ particles is multiplied by $1/z$ and thus contributes to the residue at $z = 0$. This observation allowed Dietrich, Mang, and Pradal [36] to formulate the so-called method of residues for calculating all kinds of matrix elements involving the projected state $|\Psi_N\rangle$. For example, the average HFB energy of the projected state can be written as a ratio of two residues:

$$E^{N}_{\rm HFB} = \frac{\langle\Phi|\hat{H}|\Psi_N\rangle}{\langle\Phi|\Psi_N\rangle} = \frac{\operatorname*{Res}_{z\,=\,0}\, z^{-N-1}\langle\Phi|\hat{H}|\Phi(z)\rangle}{\operatorname*{Res}_{z\,=\,0}\, z^{-N-1}\langle\Phi|\Phi(z)\rangle}. \tag{15}$$

The invariance of the projected state with respect to the integration contour can be formulated in another way; namely, one can utilize the property that an arbitrarily shifted HFB state can be equally well used to project the particle number. Indeed, for

$$|\Psi_N(z_0)\rangle \equiv \hat{P}_N|\Phi(z_0)\rangle = \frac{1}{2\pi i}\oint_C {\rm d}z\, z^{\hat{N}-N-1}|\Phi_N(z_0)\rangle \tag{16}$$

we trivially have

$$|\Psi_N(z_0)\rangle = z_0^{N}|\Psi_N\rangle, \tag{17}$$

i.e., projection from a shifted HFB state changes only the phase and normalization of the projected state. We refer to this property as *shift invariance.*

### C. HFB sum rules

Since the HFB state (11) is a superposition of projected states (13),

$$|\Phi\rangle = \sum_{N=0}^{\infty}|\Psi_N\rangle, \tag{18}$$

the HFB energy $E_{\rm HFB}$,

$$E_{\rm HFB} = \langle\Phi|\hat{H}|\Phi\rangle, \tag{19}$$

can be expressed as the sum of projected energies (15),

$$E_{\rm HFB} = \sum_{N=0}^{\infty}\langle\Psi_N|\Psi_N\rangle E^{N}_{\rm HFB}, \tag{20}$$

weighted by probabilities $\langle\Psi_N|\Psi_N\rangle$=$\langle\Phi|\Psi_N\rangle$=$\langle\Phi|\hat{P}_N|\Phi\rangle$ of finding a given PN component in the HFB state. Expression (20) constitutes a useful sum-rule condition, which has to be obeyed by any Hamiltonian-based HFB+PNP approach, and can be used to test the numerical precision of PNP techniques.

A similar sum rule holds for any shifted state

$$|\Phi(z_0)\rangle = \sum_{N=0}^{\infty}|\Psi_N(z_0)\rangle, \tag{21}$$

i.e.,

$$\langle\Phi(z_0)|\hat{H}|\Phi(z_0)\rangle = \sum_{N=0}^{\infty}|z_0|^{2N}\langle\Psi_N|\Psi_N\rangle E^{N}_{\rm HFB}, \tag{22}$$

where the average energy of the shifted and unnormalized HFB state is related to its HFB energy $E_{\rm HFB}(z_0)$ as

$$E_{\rm HFB}(z_0) = \frac{\langle\Phi(z_0)|\hat{H}|\Phi(z_0)\rangle}{\langle\Phi(z_0)|\Phi(z_0)\rangle}. \tag{23}$$

Finally, the sum rule for the non-diagonal matrix elements can be written as:

$$\langle\Phi|\hat{H}|\Phi(z_0)\rangle = \sum_{N=0}^{\infty}z_0^{N}\langle\Psi_N|\Psi_N\rangle E^{N}_{\rm HFB}. \tag{24}$$

### D. Transition matrix elements and transition densities

Calculation of the matrix elements in Eq. (15) between the original and shifted HFB states is straightforward, because the shifted states also belong to the family of the HFB states. In particular, their overlap is given by the Onishi formula [1], which in the canonical basis reduces to a simple expression,

$$\langle\Phi|\Phi(z)\rangle = \prod_{n>0}\left(u_n^2 + z^2 v_n^2\right). \tag{25}$$

Similarly, the generalized Wick's theorem [1] can be used for evaluation of Hamiltonian matrix elements,

$$\langle\Phi|\hat{H}|\Phi(z)\rangle = \langle\Phi|\Phi(z)\rangle\, E_{\rm HFB}(\rho_z,\chi_z,\bar{\chi}_z), \tag{26}$$

where the so-called HFB transition energy density $E_{\rm HFB}(\rho_z,\chi_z,\bar{\chi}_z)$ is a function of the shifted particle and pairing transition density matrices,

$$\begin{aligned}
\rho_z(\mathbf{r}\sigma,\mathbf{r}'\sigma') &= \langle\Phi|a^+_{\mathbf{r}'\sigma'}a_{\mathbf{r}\sigma}|\Phi(z)\rangle/\langle\Phi|\Phi(z)\rangle \\
&= \sum_n \frac{z^2v_n^2}{u_n^2+z^2v_n^2}\,\varphi_n(\mathbf{r}\sigma)\varphi_n^*(\mathbf{r}'\sigma'), \\
\chi_z(\mathbf{r}\sigma,\mathbf{r}'\sigma') &= \langle\Phi|a_{\mathbf{r}'\sigma'}a_{\mathbf{r}\sigma}|\Phi(z)\rangle/\langle\Phi|\Phi(z)\rangle \\
&= \sum_n \frac{z^2u_nv_n}{u_n^2+z^2v_n^2}\,\varphi_n(\mathbf{r}\sigma)2\sigma'\varphi_n^*(\mathbf{r}',-\sigma'), \\
\bar{\chi}_z(\mathbf{r}\sigma,\mathbf{r}'\sigma') &= \langle\Phi|a^+_{\mathbf{r}\sigma}a^+_{\mathbf{r}'\sigma'}|\Phi(z)\rangle/\langle\Phi|\Phi(z)\rangle \\
&= \sum_n \frac{u_nv_n}{u_n^2+z^2v_n^2}\,\varphi_n^*(\mathbf{r}\sigma)2\sigma'\varphi_n(\mathbf{r}',-\sigma').
\end{aligned} \tag{27}$$

The transition density matrices become the standard density matrices in the limit of $z\to 1$. For simplicity, we do not explicitly show the isospin variables; this is not essential in the context of the present work. (See Ref. [13] for a complete formulation.)

### E. Poles of transition densities

It is seen immediately from Eq. (27) that the transition density matrices have imaginary axis poles at

$$z_n = \pm i|u_n/v_n|, \tag{28}$$

and, therefore, are not analytical. These poles carry over to the HFB transition energy density as well. The poles appear beyond the origin, $z_n\neq 0$, provided all amplitudes $u_n$ are nonzero; we assume this hereafter, i.e., none of the canonical states is being blocked. We can also safely assume that all amplitudes $v_n$ are nonzero, because otherwise the corresponding states would not contribute to the density matrices at all. Of course, if there exist poles in the HFB transition energy density, they must be cancelled by the norm overlap $\langle\Phi|\Phi(z)\rangle$, because the Hamiltonian matrix element $\langle\Phi|\hat{H}|\Phi(z)\rangle$ is an analytical function of $z$.

However, as we discuss in the next section, whenever the transition energy density is not related to a Hamiltonian, or some approximations are involved in Hamiltonian's construction, the presence of the poles (28) requires special attention. For example, the exact HFB transition energy density,

$$E_{\rm HFB}(\rho_z,\chi_z,\bar{\chi}_z) = E_{\rm kin}(\tau_z) + E_{\rm field}(\rho_z) + E_{\rm pair}(\chi_z,\bar{\chi}_z), \tag{29}$$

is often split into the kinetic term $E_{\rm kin}(\tau_z)$ that depends on the kinetic transition density, the mean-field term $E_{\rm field}$ that depends on the particle transition density, and the pairing term $E_{\rm pair}$ that depends on the pairing transition densities. It was first realized in Ref. [30], and then discussed by several authors [31, 32, 37], that the poles are not cancelled separately in $E_{\rm field}$ and $E_{\rm pair}$, but only in the sum thereof, i.e., for the total HFB energy calculated for a given Hamiltonian.

As the origin of the pairing interaction is believed to be different from that of the effective interaction in the particle-hole direction, it is customary to employ *different* Hamiltonians to calculate $E_{\rm field}$ and $E_{\rm pair}$. This, however, leads to a non-analytical behavior of $E_{\rm HFB}$ due to the presence of poles in the complex $z$-plane; hence, to *a priori* contour-dependent projected HFB energies. We discuss this question in the next section in the more general context of the DFT energy functional.

## III. PARTICLE-NUMBER-PROJECTED DFT

According to the DFT, the energy density of the system, $E_{\rm DFT}(\rho,\chi,\chi^*)$, can be written as a function of the local particle $\rho(\mathbf{r})$ and pairing $\chi(\mathbf{r})$ densities obtained as the diagonal elements of the corresponding density matrices:

$$\begin{aligned}
\rho(\mathbf{r}) &\equiv \textstyle\sum_\sigma \rho(\mathbf{r}\sigma,\mathbf{r}\sigma) = \sum_{n\sigma} v_n^2|\varphi_n(\mathbf{r}\sigma)|^2 \\
\chi(\mathbf{r}) &\equiv \textstyle\sum_\sigma(-2\sigma)\chi(\mathbf{r}\sigma,\mathbf{r},-\sigma) = \sum_{n\sigma} u_nv_n|\varphi_n(\mathbf{r}\sigma)|^2.
\end{aligned} \tag{30}$$

The nuclear density functionals for time-even systems also depend on kinetic $\tau$ and spin-orbit $\mathsf{J}$ densities. An even larger set of densities enters the energy density for time-odd systems [13, 38]. For simplicity, we discuss here the dependence on the particle density only, because extension to other densities is straightforward.

We note in passing that the densities corresponding to the shifted HFB state $|\Phi(z)\rangle$ (6) can be written as:

$$\begin{aligned}
\rho^z(\mathbf{r}) &= \sum_n \frac{|z|^4v_n^2}{u_n^2+|z|^4v_n^2}\,\sum_\sigma|\varphi_n(\mathbf{r}\sigma)|^2, \\
\chi^z(\mathbf{r}) &= \sum_n \frac{z^2u_nv_n}{u_n^2+|z|^4v_n^2}\,\sum_\sigma|\varphi_n(\mathbf{r}\sigma)|^2.
\end{aligned} \tag{31}$$

### A. Transition energy density

In the DFT approach, the Hamiltonian of the system does not appear explicitly; hence, the projected energy cannot be calculated as its expectation value in the projected state. However, since the DFT energy density is most often postulated, not derived, we can apply the same philosophy to the projected energy, i.e., we can postulate the projected functional. In doing so, we have to guarantee that it reverts to the projected HFB energy (15) when the system is described by a Hamiltonian. In the present study, we do not discuss the construction of the projected DFT functional, but simply assume, as in most calculations up to now, that the DFT transition energy density $E_{\rm DFT}(\rho_z,\chi_z,\bar\chi_z)$ is the same as the DFT energy density $E_{\rm DFT}(\rho,\chi,\chi^*)$, but with densities $\rho$, $\chi$, and $\chi^*$ (30) replaced by the transition densities $\rho_z$, $\chi_z$, and $\bar\chi_z$ (27). This guarantees that in the limit of $z\to 1$, the projected functional gets back to the usual form.

Since the overlap (25) and HFB transition energy density (26) depend only on the shift parameter $z$ and not on its complex conjugation $z^*$, it is natural to restrict further considerations to the DFT transition energy density parametrized in the same way, i.e.,

$$E^*_{\rm DFT}(z) = E_{\rm DFT}(z^*). \tag{32}$$

Moreover, by construction, the DFT transition energy density depends only on $z^2$, and therefore it must be a symmetric function of $z$,

$$E_{\rm DFT}(-z) = E_{\rm DFT}(z). \tag{33}$$

### B. Projected DFT energy

Based on the above discussion, we postulate the projected DFT energy in the form:

$$E^N_{\rm DFT} = \frac{\oint_C dz z^{-N-1}\langle\Phi|\Phi(z)\rangle E_{\rm DFT}(\rho_z,\chi_z,\bar\chi_z)}{2\pi i \operatorname*{Res}\limits_{z=0} z^{-N-1}\langle\Phi|\Phi(z)\rangle}. \tag{34}$$

At variance with the Hamiltonian-based HFB theory, the projected DFT energy may depend on the integration contour $C$. Moreover, the numerator in Eq. (34) is, in general, not equal to the residue at $z$=0 as in Eq. (15). Consequently, both the transition energy density *and* the contour $C$ define the projected energy in DFT. Since the projected DFT energy (34) must be real, in view of condition (32), we restrict further considerations only to contours which are symmetric with respect to the real $z$-axis. Accordingly, only the upper-half contour $C_+$ above the real axis can be considered and $\oint_C dz\cdots = 2{\rm Re}[\oint_{C_+} dz\cdots]$.

### C. Analytic properties

In order to proceed further, we must investigate the analytic structure of the integrand $\mathcal{E}_N(z)$ appearing in the numerator of Eq. (34):

$$\mathcal{E}_N(z) = z^{-N-1}\langle\Phi|\Phi(z)\rangle E_{\rm DFT}(\rho_z,\chi_z,\bar\chi_z). \tag{35}$$

Let us first discuss the case when the DFT energy density $E_{\rm DFT}(\rho,\chi,\chi^*)$ is a polynomial in local densities; hence, the DFT transition energy density $E_{\rm DFT}(\rho_z,\chi_z,\bar\chi_z)$ is a polynomial in transition densities. The case of fractional power dependence requires special attention and will be discussed in Sec. III F.

Within the polynomial assumption, poles of the transition densities (28) do or do not appear as poles of the integrand (35), depending on the structure of the DFT transition energy density. [For instance, quadratic ($p$=2) and cubic ($p$=3) terms are characteristic of two-body and three-body interactions, respectively.] On the one hand, each polynomial term of the order $p$ in densities (30) brings about a pole of the order $p$. On the other hand, each term in the overlap (28) produces a zero of the order $q$, where $q$ is the degeneracy factor of the HFB density matrix with the two-fold Kramers degeneracy not counted. (Note that the product in Eq. (28) contains only one term for each canonical pair.) In particular, for the spherical shell of angular momentum $j$, the degeneracy is $q = j+\frac{1}{2}$.

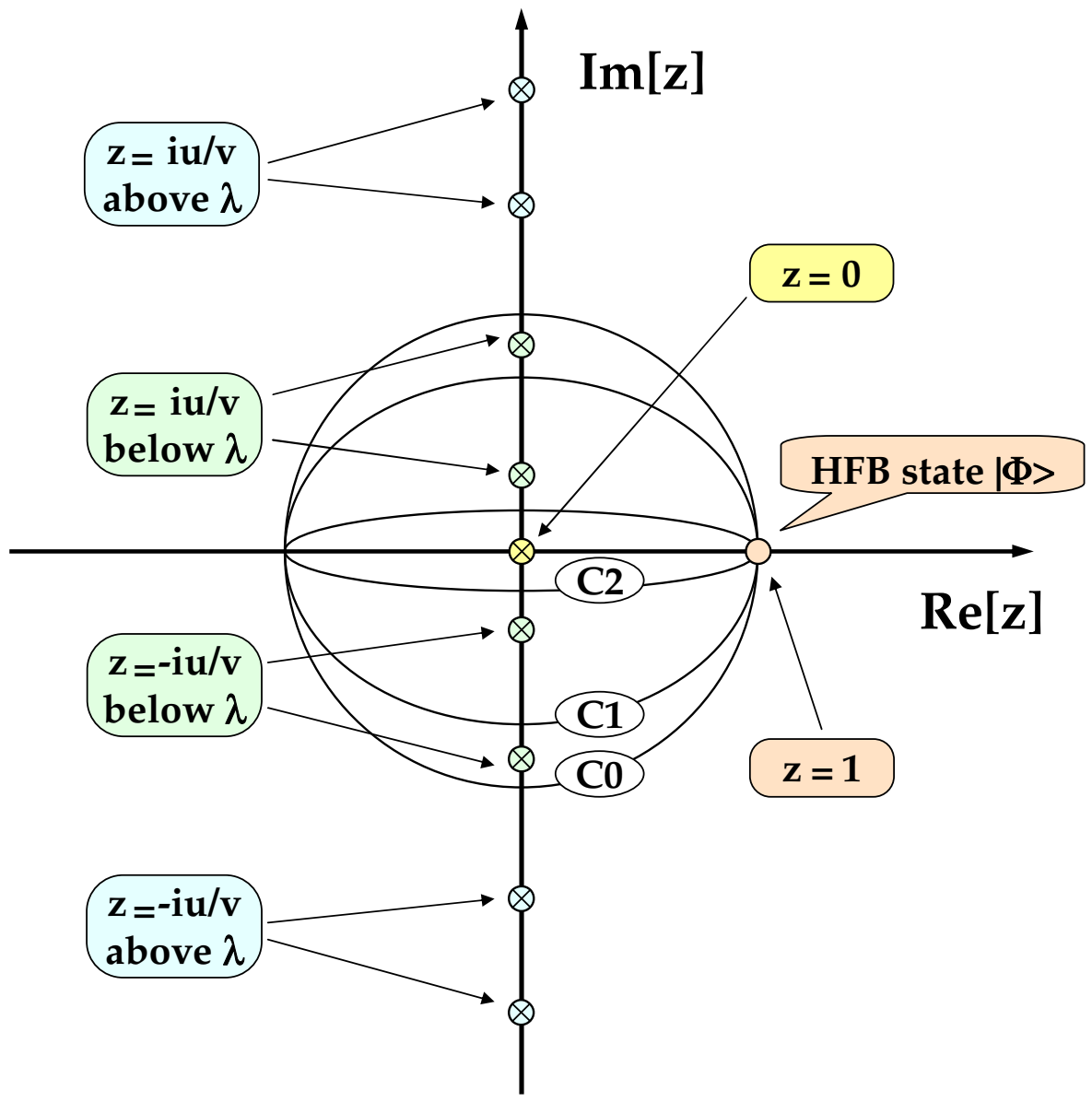

FIG. 1: (color online) Schematic illustration of the analytic structure of the integrand (35) in the complex $z$-plane (see text). Small crossed circles denote imaginary axis poles. Three integration contours ($C0$, $C1$, $C2$), symmetric with respect to the real axis, are indicated. The poles having particle character (corresponding to canonical states lying above the chemical potential $\lambda$) are located outside the unit circle $C0$ while the hole poles lie inside $C0$. The unprojected ground state wave function corresponds to $z$=1.

When the poles of transition densities and zeros of the overlap $\langle\Phi|\Phi(z)\rangle$ are combined, the poles in $\mathcal{E}_N(z)$ are of the order $p-q$. For single-particle states that have only two-fold Kramers degeneracy ($q$=1), and for the terms with $p$=2, one obtains the first-order poles in $\mathcal{E}_N(z)$ with, in general, non-zero residues. Non-vanishing residues may also appear for higher-order poles corresponding to terms with $p>2$. On the other hand, for four-fold degenerate states with $q$=2, terms with $p$=2 do not produce poles in $\mathcal{E}_N(z)$, and only terms with $p>2$ may give rise to poles with non-vanishing residues. As discussed in detail in Ref. [30], for the energy density derived from a Hamiltonian, additional cancellations between terms originating from particle-hole and pairing channels occur, and the first-order poles disappear.

In Fig. 1 we schematically illustrate the analytic struc-ture of the integrand (35). Crossed circles on the imaginary axis represent poles of $\mathcal{E}_N(z)$. Apart from the pole at $z = 0$, the integrand may have poles (28) distributed symmetrically in pairs with respect to the real axis. Poles located within the unit circle (C0 in Fig. 1) correspond to the canonical states with occupation numbers larger than 0.5, or with $u_n/v_n \le 1$, i.e., with canonical energies below the Fermi energy $\lambda$. Similarly, poles outside the unit circle correspond to canonical states lying above the Fermi energy.

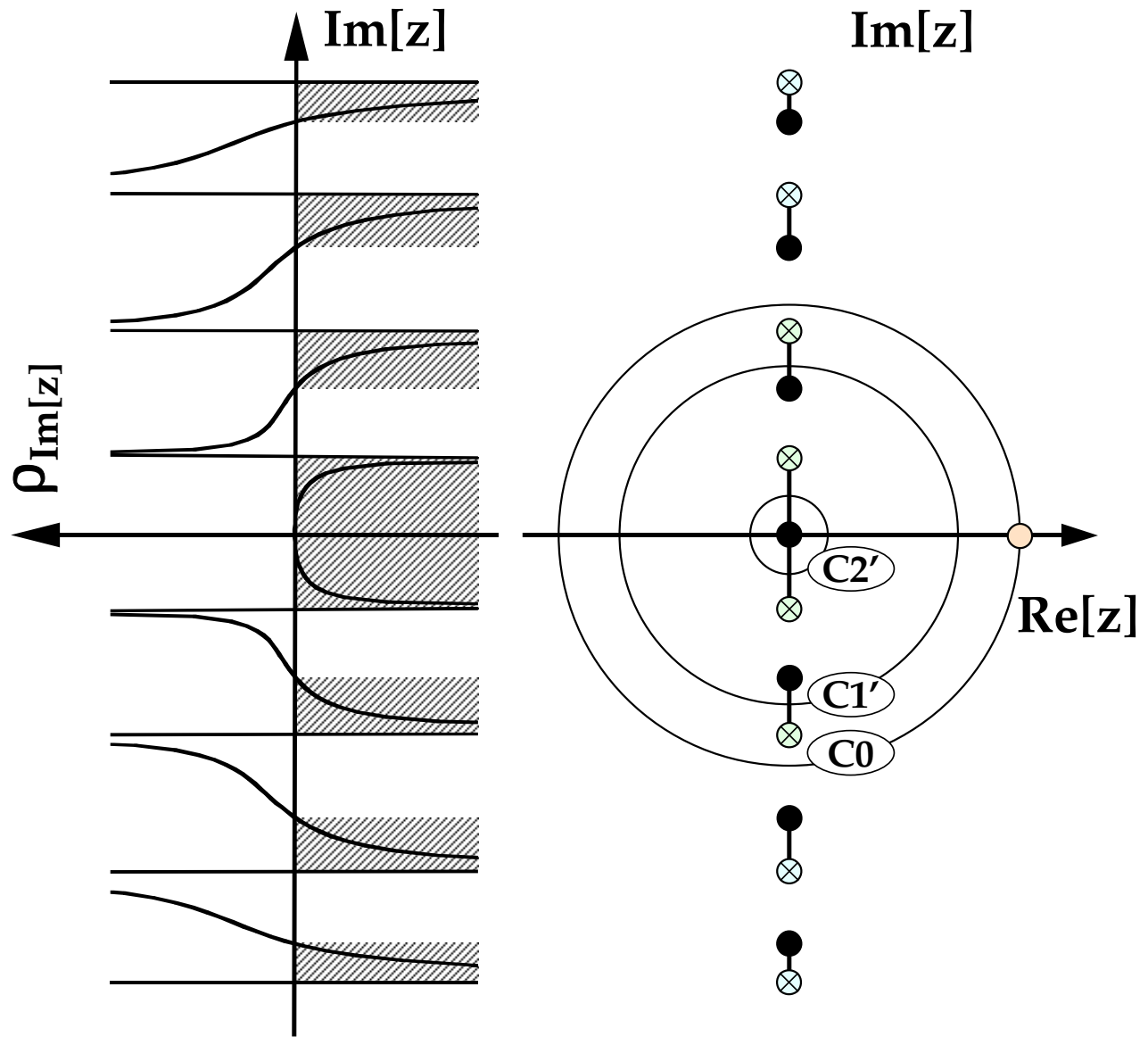

FIG. 2: Schematic illustration of analytic structure of the particle transition density at a fixed point in space **r**. The poles (crossed circles) and zeros (dots) of $\rho_z$ are located on the imaginary-$z$ axis. The regions of real negative $\rho_z$ are shaded. Three circular integration contours $C0$, $C1'$, and $C2'$ are indicated. See text for details.

The unprojected HFB ground state $|\Phi\rangle$, located at $z$=1, is shifted along the integration contour $C$, and its overlap and DFT transition energy contribute to the integrand of Eq. (35). Standard projection formula (1) corresponds to the unit circle $C0$. Contours $C1$ and $C2$ encircle a fewer number of poles in $\mathcal{E}_N(z)$, with contour $C2$ surrounding only the single pole at the origin. Shapes of these contours are irrelevant, and only the points at which they cross the imaginary axis matter. For example, contours $C1$ and $C2$, shown in Fig. 1, are equivalent to circular contours $C1'$ and $C2'$ of Fig. 2, the latter being more practical in calculations. If the residues of the poles inside the unit circle are non-zero, the three integration contours shown in Fig. 1 may give different projected energies. Of course, contours including poles located outside the unit circle (not shown in Fig. 1) may still give different results.

### D. Residues

Let us now discuss the residues of the integrand (35). From Eqs. (25) and (33), we see that the integrand is an odd function of $z$,

$$\mathcal{E}_N(-z) = -\mathcal{E}_N(z). \tag{36}$$

This is obvious for even particle numbers $N$, for which Eq. (25) has been derived, while for odd $N$, an additional power of $z$ appears when shifting the blocked HFB state,

$$|\Phi^{\rm odd}\rangle = a^+_{n_0} \prod_{n\neq n_0>0} \left(u_n + v_n a^+_n a^+_{\bar n}\right) |0\rangle, \tag{37}$$

i.e.,

$$|\Phi^{\rm odd}_N(z)\rangle = z a^+_{n_0} \prod_{n\neq n_0>0} \left(u_n + z^2\, v_n a^+_n a^+_{\bar n}\right) |0\rangle, \tag{38}$$

which gives

$$\langle\Phi^{\rm odd}|\Phi^{\rm odd}_N(z)\rangle = z \prod_{n\neq n_0>0} \left(u_n^2 + z^2 v_n^2\right), \tag{39}$$

and renders the integrand (35) an odd function of $z$ also for odd systems.

Near the pole (28), the term in the integrand that produces the residue has the structure:

$$\mathcal{E}_N(z) \simeq \frac{\mathcal{R}_n(z)}{u_n^2 + z^2 v_n^2}, \tag{40}$$

where $\mathcal{R}_n(z)$ is an odd function of $z$, regular at the pole. Similarly, for the pole at $z$=0 we have

$$\mathcal{E}_N(z) \simeq \frac{\mathcal{R}_0(z)}{z}. \tag{41}$$

Therefore, for pairs of poles that are symmetric with respect to $z$=0, the residues,

$$\begin{aligned} \underset{z\,=\,\pm i|u_n/v_n|}{\rm Res} \mathcal{E}_N(z) &= \lim_{z\to\pm i|u_n/v_n|} \frac{(z\mp i|u_n/v_n|)\mathcal{R}_n(z)}{u_n^2+z^2v_n^2} \\ &= \frac{\mathcal{R}_n(i|u_n/v_n|)}{2i|u_nv_n|}, \end{aligned} \tag{42}$$

have identical values. Hence, poles below and above the real axis yield the same contribution to the contour integral. Based on this consideration, the projected DFT energy (34), expressed in terms of residues, reads:

$$E^N_{\rm DFT} = \frac{\mathcal{R}_0(0) + 2\sum_{n\in C} \frac{\mathcal{R}_n(i|u_n/v_n|)}{2i|u_nv_n|}}{\underset{z\,=\,0}{\rm Res}\,\langle\Phi|\Phi(z)\rangle} \tag{43}$$

or

$$E^N_{\rm DFT} = \sum_{n=0}^{\bar n} E^N_{\rm DFT}(n), \tag{44}$$

where $E^N_{\rm DFT}(n)$ denotes the contribution from the $n$th pole, including the $n$=0 pole at the origin up to $n = \bar n$ (last pole encircled by $C$).

As an example, we explicitly calculate the residues for a term that depends on the squared particle density,

$$E_{\rm DFT}(\rho_z) = C^\rho \int {\rm d}^3\mathbf{r}\rho_z^2(\mathbf{r}), \tag{45}$$

with

$$\rho_z(\mathbf{r}) = \sum_n \frac{z^2 v_n^2}{u_n^2 + z^2 v_n^2} \sum_\sigma |\varphi_n(\mathbf{r}\sigma)|^2 \tag{46}$$

and $C^\rho$ being a coupling constant. Assuming a two-fold Kramers degeneracy, the corresponding residue at $\pm i|u_n/v_n|$ is:

$$\operatorname*{Res}_{z\,=\,\pm i|u_n/v_n|} \mathcal{E}_N(z) = 2C^\rho v_n^2 \left(-\frac{v_n^2}{u_n^2}\right)^{\frac{N-2}{2}} \int {\rm d}^3\mathbf{r} \left(\sum_\sigma |\varphi_n(\mathbf{r}\sigma)|^2\right)^2 \prod_{m\neq n>0} v_m^2 \left(\frac{u_m^2}{v_m^2} - \frac{u_n^2}{v_n^2}\right). \tag{47}$$

One can see that residues can be very large for poles corresponding to canonical states that have occupation numbers close to 1. These very large contributions to the projected DFT energy must be compensated by a similarly large contribution from the single pole at $z$=0. Therefore, within the DFT formalism, one cannot use the HFB expression (15) that involves only one residue at $z$=0.

Recall from our discussion in Sec. III C that the poles have the order of $p-q$. In the above example, the polynomial order is $p$=2; hence, the residue (47) must vanish if the degeneracy factor $q \geq 2$. This is indeed the case as for $q > 1$ $u_m^2 = u_n^2$ for at least one value of $m \neq n$.

### E. The DFT sum rules

The HFB sum rules derived in Sec. II C are based on the linearity of the Hamiltonian, by which a matrix element involving the HFB state is a sum of matrix elements calculated for all the PNP components (18). In order to derive the analogous sum rules for the projected DFT energies, one can only use properties of the underlying transition energy density. To this end, we recall that in the HFB theory, the mixing of particle numbers corresponds to the broken U(1) gauge symmetry, and that the PNP actually corresponds to expanding the HFB state in irreducible representations of this group. This observation can be extended to the DFT transition energy density, expanded in these same irreducible representations, with the projected DFT energies being the expansion coefficients. The resulting sum rules must follow from the closure relations on the group manifold.

These general remarks can be expressed in an explicit form in the following way. By using integration contours that are circles of radius $|z_0|$ around the origin, $z = z_0 e^{i\phi}$, we have the following expression for the projected DFT energy (34)

$$\langle\Psi_N|\Psi_N\rangle E^N_{\rm DFT} = \frac{z_0^{-N}}{2\pi}\int_0^{2\pi} {\rm d}\phi e^{-iN\phi}\mathcal{E}(\phi), \tag{48}$$

where by $\mathcal{E}(\phi)$ we denoted the part of the integrand that does not depend on $N$, i.e.,

$$\mathcal{E}(\phi) = \langle\Phi|\Phi(z)\rangle E_{\rm DFT}(\rho_z,\chi_z,\bar{\chi}_z) \quad \text{at} \quad z = z_0 e^{i\phi}. \tag{49}$$

Hence, the DFT projected energy is given by a Fourier transform of $\mathcal{E}(\phi)$. Since the Fourier components constitute a complete set of functions on a circle, $\sum_{N=0}^\infty e^{-iN\phi} = 2\pi\delta(0)$, we obtain the DFT sum rule,

$$\langle\Phi|\Phi(z_0)\rangle E_{\rm DFT}(\rho_{z_0},\chi_{z_0},\bar{\chi}_{z_0}) = \sum_{N=0}^\infty z_0^N \langle\Psi_N|\Psi_N\rangle E^N_{\rm DFT}, \tag{50}$$

which is the analogue of the HFB sum rule for matrix elements (24). For $z_0$=1, we obtain the DFT counterpart of the HFB sum rule (20):

$$E_{\rm DFT}(\rho,\chi,\chi^*) = \sum_{N=0}^\infty \langle\Psi_N|\Psi_N\rangle E^N_{\rm DFT}. \tag{51}$$

We note that in the above derivations, $z_0$ is an arbitrary complex number; its modulus fixes the radius of integration contour, while its phase gives the point on the circle that fixes the starting point of the integral in Eq. (48). This starting point has obviously no importance for the value of the integral. The sum rule (50) gives, therefore, a representation of the DFT transition energy density in terms of a series expansion in $z_0$, which converges only on the ring between the poles. For each such ring, the projected DFT energies $E^N_{\rm DFT}$ are different, and the DFT transition energy density is thus equal to a different series expansion. It is obvious that these different values of the projected DFT energies do not contradict the continuity of the DFT transition energy density. In this way, all projected DFT energies for arbitrarily chosen contours of integration correspond to this same common DFT energy functional.

## F. Density-dependent terms with fractional powers

Let us now analyze the terms in the DFT energy density that depend on fractional powers $\alpha$ of the local density. In many functionals related to the Skyrme interaction, and for the Gogny force, such terms are quite often postulated, both in the particle-hole and pairing channels (see Ref. [12] for a review). In particular, the familiar density-dependent term of the Skyrme force, which is proportional to $\rho^\gamma(\mathbf{r})$, produces a contribution of the order of $\alpha = 2+\gamma$ to the DFT energy density. Similarly, the density-dependent, zero-range term of the Gogny force yields a contribution to the DFT energy density that is of $\alpha = 1+\gamma$ order, provided the particle-hole and pairing terms are consistently added. By taking into account the degeneracy factors $q$ discussed in Sec. III C, the resulting poles are of the order of $\alpha - p = 2+\gamma-p$ and $\alpha - p = 1+\gamma-p$, respectively. Since typical values of $\gamma$ are between 0 and 1, the DFT transition energy density always has poles at $z_n = \pm i|u_n/v_n|$ for non-degenerate states ($q = 1$). But more importantly, the fractional powers lead to the *multivalued* DFT transition energy density on the complex-$z$ plane.

In the standard treatment of fractional powers $\alpha$ of a complex function, cuts along the negative real axis must be introduced. In order to apply this procedure to fractional powers of the local transition density (46), we must identify on the complex $z$ plane the lines along which $\rho_z(\mathbf{r})$ is real negative.

Obviously, $\rho_z(\mathbf{r})$ is real positive along the real $z$ axis and real along the imaginary $z$ axis. In order to simplify the discussion, let us assume that the sum in Eq. (46) is finite, which is always the case in any practical calculation. In such a case, $\rho_z(\mathbf{r})$ has a finite number, say $M$, of different first-order poles along the positive imaginary axis, and the same number $M$ of poles along the negative imaginary axis. Moreover, since all coefficients in Eq. (46) are positive, $\rho_z(\mathbf{r})$ must have a first-order zero between each pair of poles on the positive imaginary axis, and similarly on the negative imaginary axis. Since $\rho_z(\mathbf{r})$ also has a second-order zero at $z = 0$, we conclude that it has altogether $2M$ zeros on the imaginary axis. It is also obvious that $\rho_z(\mathbf{r})$ is a rational function with a $2M$-order polynomial in the numerator, and thus we conclude that all the zeros of $\rho_z(\mathbf{r})$ are located on the imaginary axis. Therefore, the cuts for possible fractional powers $\alpha$ must be located along the imaginary axis, and connect zeros of $\rho_z(\mathbf{r})$ with its adjacent poles.

The above discussion is visualized in Fig. 2. The left portion shows schematically the transition density $\rho_{\mathrm{Im}[z]}(\mathbf{r})$ along the imaginary axis Im$[z]$ oriented vertically. The plot illustrates the transition density (46) in one selected point of space $\mathbf{r}$, i.e., values of wave functions at $\mathbf{r}$ enter only as numerical coefficients. There appear four poles and three zeros of $\rho_{\mathrm{Im}[z]}$ on the positive imaginary axis, the same number of poles and zeros on the negative imaginary axis, and the second-order zero at the origin. Sections of the imaginary axis where the density is negative are shaded. In the right portion of Fig. 2 we show poles (crossed circles) and zeros (full dots) of the transition density on the complex $z$ plane, along with the three integration contours $C0$, $C1'$, and $C2'$ discussed above. The cuts in the complex $z$-plane connecting zeros and poles, corresponding to real negative values of $\rho_z$, are indicated by vertical segments.

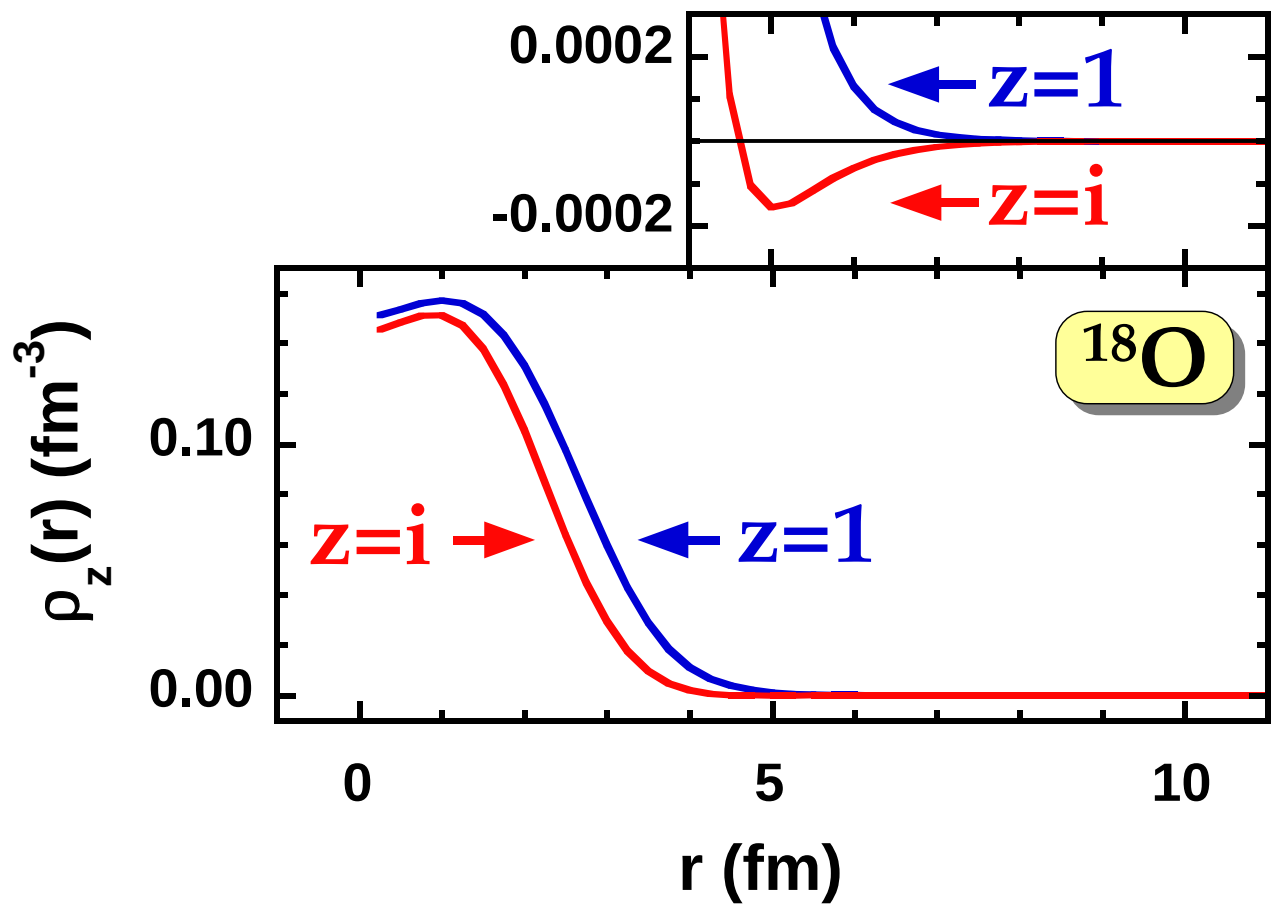

FIG. 3: (color online) Ordinary ($z = 1$) and transition ($z = i$) densities in $^{18}$O calculated in HFB+SLy4 as functions of $r$. The upper part of the Figure shows (in extended scale) the small region of radii where the transition density becomes negative.

While the location of the poles is independent of $\mathbf{r}$, the position of zeros of the transition density is $\mathbf{r}$-dependent. In order to visualize this, in Fig. 3 we plot the total ordinary ($z = 1$) and transition ($z = i$) densities in $^{18}$O obtained within the SLy4 energy density functional. One can see that the transition density is positive almost everywhere; only in a very narrow region near $r \simeq 5$ fm does it become slightly negative, as shown in the upper part where the scale is expanded by a factor of 1000. Such a behavior of $\rho_z(\mathbf{r})$ at $z^2 = -1$ can be easily understood from Eq. (46). Indeed, strongly occupied states with $v_n^2 \simeq 1$ always yield positive contributions, while negative contributions of states with $v_n^2 < u_n^2$ can only appear in the surface region where the least bound canonical states dominate.

By the same token, we can see that strong negative contributions may appear when the integration contour passes slightly below a pole located just above the unit radius, i.e., $v_n^2$ is slightly smaller then $u_n^2$. Such a situation is predicted in $^{26}$O, where the occupation probability of the canonical $2d_{3/2}$ state equals 0.486. As shown in the bottom part of Fig. 4, the transition density at $z = i$ is dominated by this particular contribution and becomes strongly negative beyond $r \simeq 2$ fm.

We are now ready to discuss contour integration of terms depending on fractional powers of the transition density. Contour $C0$ shown in Fig. 2 crosses the imaginary axis in sections where there is no cut, and thus it always stays on the same Riemann sheet. On the other hand, contours $C1$ and $C2$ of Fig. 1, or contours $C1'$ and $C2'$ of Fig. 2, cross the imaginary axis by passing through cuts onto another Riemann sheet. Since the transition density (46) is an even function of $z$, the phase of the fractional power $\alpha$ of the transition density increases or decreases by $2\pi\alpha$ when going across *each* of the two cuts. Therefore, after returning to $z = 1$, $\rho_z(\mathbf{r})$ is multiplied by $\exp(\pm 4\pi\alpha)$, and thus it is *not* a continuous function at $z = 1$, unless $\alpha = k/2$. This is quite unacceptable as the presence of the phase creates serious problems in interpreting the projected DFT energies (34) (see, e.g., the sum rule condition discussed in Sec. III E).

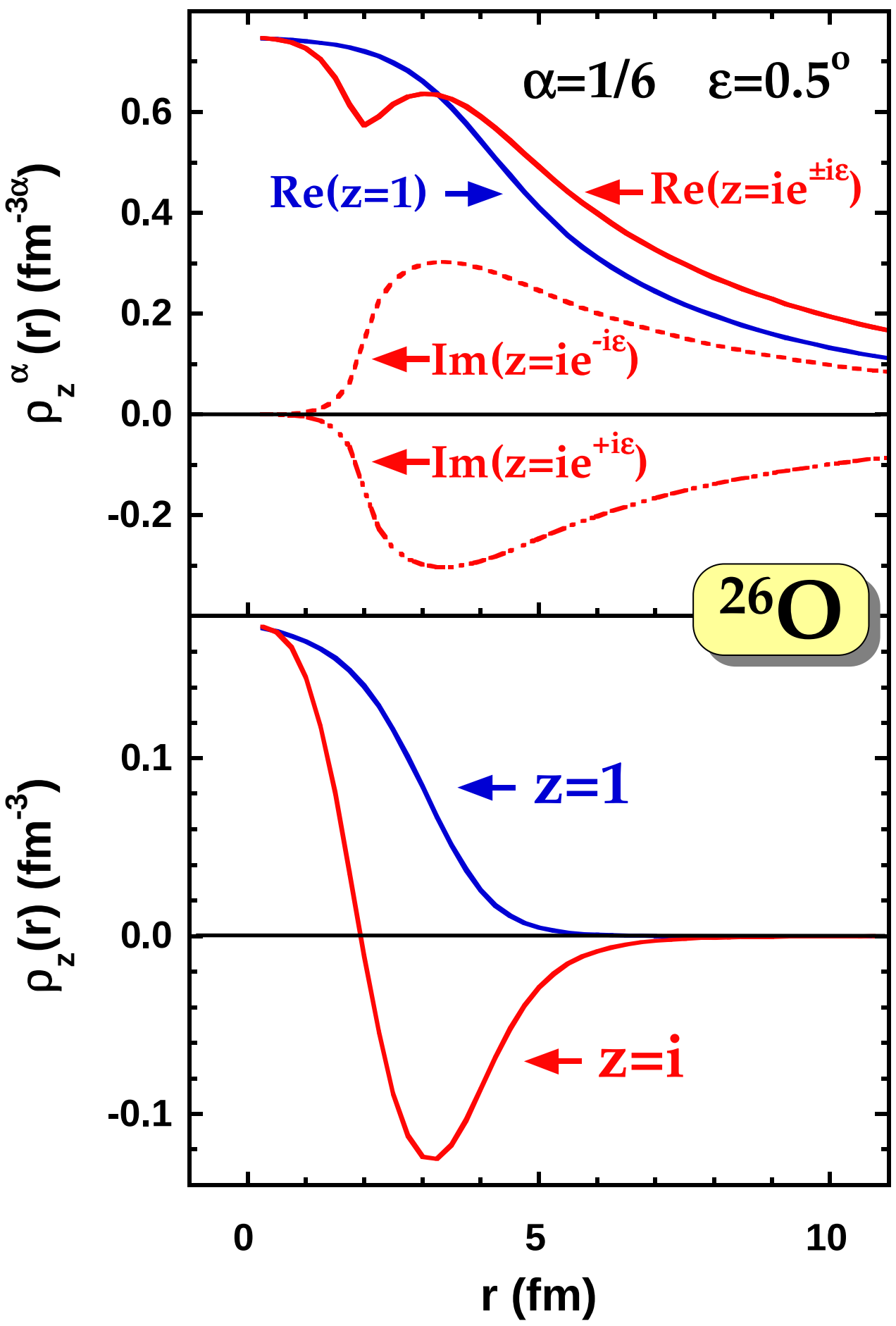

FIG. 4: (color online) Bottom: ordinary ($z = 1$) and transition ($z = i$) densities in $^{26}$O calculated in HFB+SLy4 as functions of $r$. Top: the real parts (solid lines) and imaginary parts (dashed lines) of the 1/6 powers of the ordinary ($z = 1$) and transition ($z = ie^{\pm i\epsilon}$) densities.

Formally, by using powers of square roots in density-dependent terms, i.e., $\alpha = k/2$, one can guarantee that the integration contours return onto the original Riemann sheet, and that the transition energy density is a continuous function of $z$. However, even in such a case, one important property of the DFT transition energy density (33) is lost, namely, the density-dependent term in the energy density becomes an *odd* function of $z$, and the corresponding term in the integrand (35) becomes an *even* function of $z$. This is so, because the square root has opposite signs on the two Riemann sheets in question. Consequently, contour integrals of such terms would vanish and the density-dependent terms would yield zero contribution to the PN-projected energy. This is a rather disastrous result. Hence, we are forced to conclude that the use of continuous contours for fractional powers is not a viable prescription for constructing the projected DFT energies.

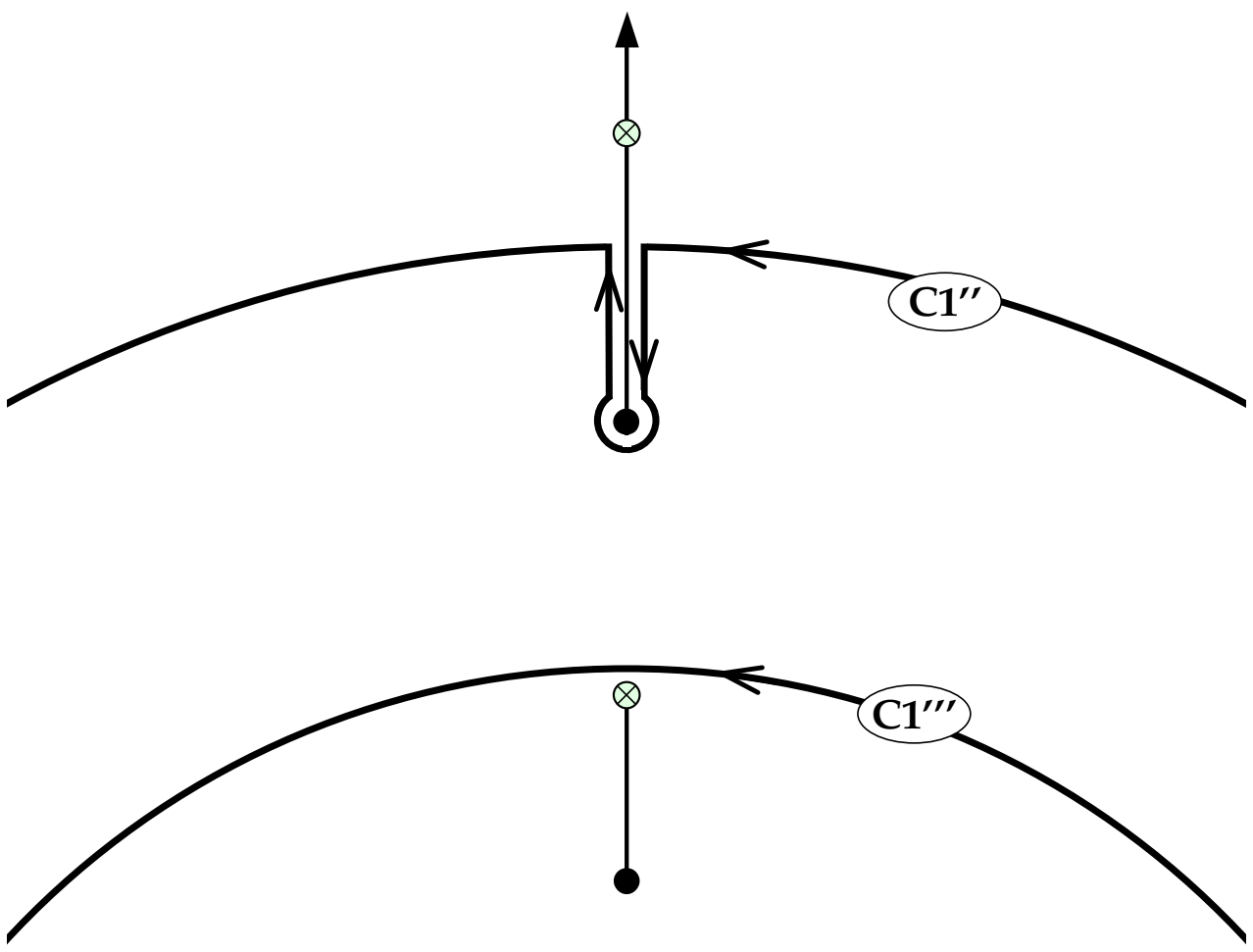

FIG. 5: Modified closed contour $C1''$ that encircles the zero of the transition density. The poles (zeros) of $\rho_z$ are marked by crossed circles) (dots). The equivalent circular contour $C1'''$ lying between the zero of $\rho_z$ and the previous pole is also indicated. See text for more details.

Let us now discuss the way of evaluating contour integrals in all practical PNP calculations up to now. Unfortunately, such calculations have always disregarded the analytic structure of the underlying integrands. In fact, the fractional powers of transition density,

$$\rho_z^\alpha(\mathbf{r}) = |\rho_z(\mathbf{r})| \exp\left\{ i \arg\left[\rho_z(\mathbf{r})\right]/\alpha \right\}, \qquad (52)$$

are practically determined by computer compilers. In Eq. (52), the so-called argument $\arg[\rho_z(\mathbf{r})]$ of $\rho_z(\mathbf{r})$ is defined as the phase of the complex variable $\rho_z(\mathbf{r})$; hence, it is contained in the interval of $-\pi \div \pi$. This usual prescription corresponds to stepping over the cut whenever the contour approaches the imaginary axis for $\arg[\rho_z(\mathbf{r})] = \pm\pi$, i.e., for real negative transition densities $\rho_z(\mathbf{r}) < 0$. In this way, the integrand is always calculated on the same Riemann sheet, but *the integration contour is not closed.*

The contour can be closed by adding a piece that goes around the zero of the transition density $\rho_z(\mathbf{r})$. This is illustrated in Fig. 5 that shows a modification of contour $C1'$ of Fig. 2 near the positive imaginary $z$-axis. [An analogous mirror-like detour is made near the negative imaginary axis.] The resulting contour $C1''$ always stays on the same Riemann sheet and, therefore, the integration result does not depend on the radius. On the other hand, the contribution due to the additional path surrounding the zero is affected by the discontinuity of the integrand along the cut. Such a discontinuity in the transition density is shown in the top panel of Fig. 4 for $\alpha$=1/6 (52). Since the ordinary density ($z = 1$) is real and positive, its fractional power is also real and positive. On the other hand, transition densities (46) corresponding to $z_\pm = ie^{\pm i\varepsilon}$ with $\varepsilon = 0.5°$, i.e., near the positive imaginary axis, are complex. While their real parts are practically identical on both sides of the cut, their imaginary parts have opposite signs; hence, a discontinuity is encountered. [Of course, since the directions of integration are opposite, the contributions to the PNP energy from both segments $z_\pm$ of the additional path are identical.]

When transforming the contour integration in Eq. (34) into the integral along the imaginary axis $y$, one must do a change of variables from $z$ to $iy$. This introduces the additional factor $i^{-N}$ in the integrand. For that reason, for even $N$, the discontinuity in the *imaginary* part of the density-dependent term of fractional order contributes to the *real* part of the projected DFT energy. As the discussion in Sec. III D proves, the same holds for odd values of $N$.

Figure 5 also shows the circular contour $C1'''$ lying between the zero of $\rho_z$ and the previous pole $|z_{n-1}|$. This contour is formally equivalent to the deformed contour $C1''$ but it is easier to handle in practical applications. The radius of $C1'''$ must be slightly greater than $|z_{n-1}|$ and smaller than the lowest zero of $\rho_z(\mathbf{r})$, minimized over the whole space $\mathbf{r}$, associated with the branching point corresponding to $z_n$. The use of contour $C1'''$ guarantees that the integration of fractional-order terms is done properly.

Altogether, blind application of prescription (52) can lead to spurious and entirely uncontrolled contributions to the projected DFT energies. Excepting Ref. [23], this fact has been entirely overlooked in all practical applications of the PNP method to date, and casts serious doubts on the reliability of the obtained results. Largest contributions are, of course, obtained when the integration contour passes slightly below a pole of the DFT transition energy density. For the Skyrme functionals in Sec. IV B, we present specific examples of such situations.

The appearance of spurious contributions is, in fact, independent of the order of divergence at the pole. Therefore, it also shows up for "integrable" poles, diverging with powers of $\alpha - p = 1 + \gamma - p < 1$, discussed for the Gogny force in Ref. [37].

## IV. NUMERICAL EXAMPLES

In order to illustrate theoretical findings presented in Sec. III, we carried out numerical calculations within the Skyrme-DFT method. We used the code HFBTHO [39] which is capable of handling spherical and axially deformed nuclei within the Lipkin-Nogami (LN) approximation followed by the PNP. This corresponds to the projection-after-variation (PAV) method of restoring the PN symmetry. By using a new version of HFBTHO, we also performed full variation-after-projection (VAP) calculations analogous to those of Ref. [23].

As illustrative examples, we study spherical and deformed configurations in $^{18}$O and in $^{32}$Mg calculated using the Skyrme functionals SIII [40] and SLy4 [41]. These two parametrizations differ in a significant way with respect to the PNP method. The density-dependent term of SIII contributes to the energy density as $(\rho_n + \rho_p)\rho_n\rho_p$. Therefore, both in the neutron and proton subsystems, the powers $p$ (Sec. III E) of the density dependence are equal to 2. Consequently, from the PNP perspective, the density-dependent term of SIII is not any different than the density-independent terms. On the contrary, the density-dependent term of SLy4 is proportional to $[\rho_n + \rho_p]^{1/6}$ and exemplifies the case of fractional-power dependence discussed in Sec. III F. The contact pairing force of the volume type (density-independent) was used in the particle-particle channel. All calculations have been performed in the spherical harmonic-oscillator basis of $N_0 = 6$ or 10 shells, for $^{18}$O or $^{32}$Mg, respectively.

### A. Numerical accuracy

To calculate residues, we take circular contour integrals of radius $r_0$:

$$z = r_0 e^{i\phi}. \tag{53}$$

The integrals are evaluated using the Fomenko discretization method [42, 43], whereby values of integrands are summed up at gauge angles $\phi_k = \frac{k\pi}{L}$ for $k = 0, \ldots, L-1$. This corresponds to the upper half circle in the complex $z$-plane and, as discussed in Sec. III B, only the real part of the integral is kept. For analytic integrands, the Fomenko method delivers exact results up to admixtures of wave functions with $N \pm L, N \pm 2L, N \pm 3L, \ldots$ particles. The main question in applying this method to non-analytic integrands, which have poles in the complex plane, is to what extend can it deliver equally accurate numerical results.

The Fomenko method clearly fails when there is a pole (28) lying just on the integration contour, $r_0 = |z_n|$, and an even number of points $L$ is used. In such a case, the integration point with $k$=$L/2$ is located exactly at the pole of the integrand. Therefore, in most practical calculations, an odd number of integration points, most often $L = 7$ or 9, was used.

However, a more stringent condition on $L$ results from the fact that the discretization method must fail whenever the integrand varies too rapidly between two neighboring integration points. Therefore, the spacing between points $\pi r_0/L$ must be appropriately smaller than the distance from the pole. For odd values of $L$, the integration points corresponding to $k = (L \pm 1)/2$ are closest to the imaginary axis; hence, one arrives at the condition

$$\frac{\pi r_0}{L} < \sqrt{(r_0 - |z_n|)^2 + \left(\frac{\pi r_0}{2L}\right)^2}, \tag{54}$$

or

$$L > \frac{\sqrt{3}\pi r_0}{\sqrt{4}|r_0 - |z_n||}. \tag{55}$$

In the present study, a large number of $L = 93$ integration points was used, which allows for calculating the contour integrals with radii $r_0$ that differ by as little as 3% from the position of the closest pole $|z_n|$.

### B. Dependence of projected energy on integration contours in spherical nuclei

Table I displays the results of PNP calculations performed for $^{18}$O by using circular integration contours (53) of different radii. The precision of numerical integrations was confirmed by calculating contributions from individual poles. This was done by carrying out contour integrals over small circles surrounding the poles. In this way, we determined residues from the individual poles $E^N_{\rm DFT}(n)$ and checked that their sums, $\sum_{m=0}^{n} E^N_{\rm DFT}(m)$, agree very well with the results of contour integrals along circular contours $C$, as required by the Cauchy theorem (44).

As seen in Table I, contributions of the $n = 0$ poles at $z = 0$ are huge. Therefore, the DFT residues at $z = 0$ cannot at all be interpreted as the projected energies, as was the case for the PNP HFB theory, Eq. (15). Residues at $z = 0$ are cancelled, to a large extent, by contributions from the $1s_{1/2}$ deep-hole states, which are large because they contain large factors of the type $\left(-v_n^2/u_n^2\right)^N$ for $u_n^2 \simeq 0$ [see Eq. (47]. Contributions from other poles are also quite large, and apart from the integration contour at $|z| = 1$, none of the other contours reproduce the correct projected energy shown by a boxed number.

For the SIII parametrization, one can see that contributions from poles associated with spherical states with $j \geq 3/2$ ($q \geq 2$) are indeed equal to zero, cf. discussion in Sec. III C. This property does not hold for SLy4, for which the projected DFT energies have jumps also when the integration contours cross the $j \geq 3/2$ poles. In this case, the jumps are not related to non-zero residues, but, as discussed in Sec. III F, they are caused by the fact that the integration contours are not closed for the fractional-power terms.

Figure 6 shows the projected DFT-SLy4 energies obtained by using circular integration contours of different radii $r_0$. These calculations illustrate properties of poles listed in Table I. The contributions originating from the density-independent and density-dependent terms of the Skyrme force are separated. The latter terms yield the fractional-power terms in the DFT energy density discussed in Sec. III F. As in the SIII case, the density-independent terms exhibit jumps only at the two $j = 1/2$ poles. On the other hand, the density-dependent terms show jumps at all poles, and these jumps carry over to the total projected DFT energies shown in Table I. [The small jump at the $1d_{5/2}$ pole, 120 keV, is practically invisible in the scale of Fig. 6.] Moreover, contributions of the density-dependent terms are not constant between the poles, as would be required by the Cauchy theorem. This is caused by the prescription (52) to step over the cuts in the complex plane, and illustrates spurious contributions to the projected DFT energies discussed in Sec. III F. As shown in the blown-up inset in Fig. 6(b), these spurious contributions appear just below the pole thresholds (i.e., for small negative values of $r_0 - |z_n|$), and they can be quite large – of the order of several tens of MeV. The gradual development of spurious contributions below threshold has been explained in Sec. III F. Namely, if the contour radius is only slightly greater than $|z_{n-1}|$, the branching point associated with the pole $z_n$ is always outside for all values of $\mathbf{r}$. With increasing $r_0$, more and more branching points corresponding to different regions of space fall inside the contour, leading to the spurious behavior. As discussed earlier, one can eliminate this subthreshold effect by taking equivalent contours discussed in the context of Fig. 5. Such a procedure is illustrated by a dotted line in the inset of Fig. 6(b).

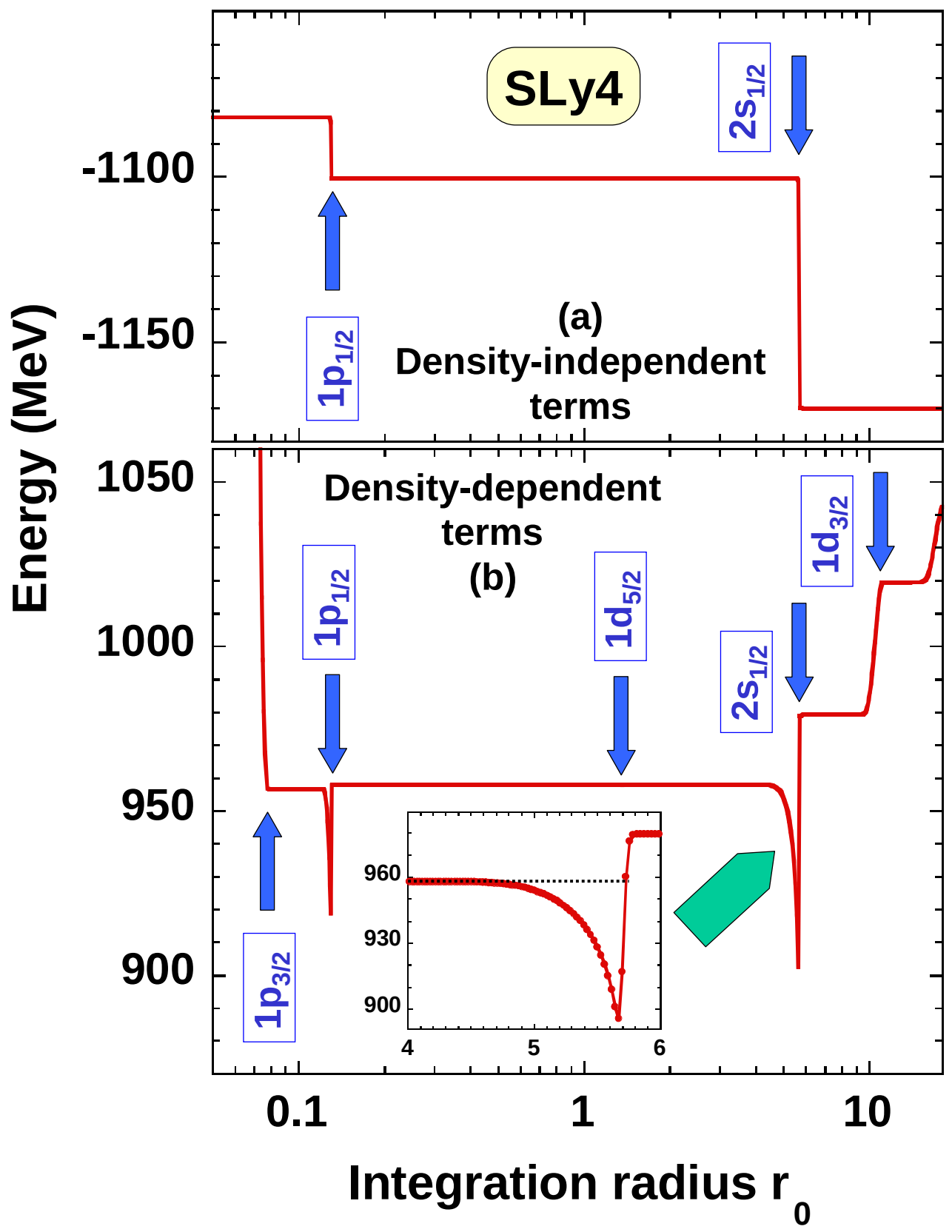

FIG. 6: (color online) The $N = 10$ projected DFT-SLy4 energies (44) calculated in $^{18}$O as functions of the integration radius $r_0$. Upper (a) and lower (b) panels show, respectively, results for the terms originating from density-independent and density-dependent parts of the Skyrme force. Positions of individual neutron poles are marked by arrows. The inset shows the results near the $2s_{1/2}$ pole on an expanded scale. The result of calculations obtained with the *equivalent* contours passing below the branching point associated with the $2s_{1/2}$ pole (cf. Fig. 5) is shown by a dotted line.

TABLE I: Contributions $E^N_{\rm DFT}(n)$ in $^{18}$O from the individual neutron poles to the projected DFT energy (44) for $N = 10$ calculated using the SIII and SLy4 Skyrme functionals. For each parametrization, the last column shows the sum of the $n$th lowest contributions (44), with the values of $E^N_{\rm DFT}$ marked by boxed numbers. Canonical energies $\epsilon_n$ and pole positions $z_n$ are also given. All energies are in MeV.

| | | SIII | | | | SLy4 | | | |
|---|---|---|---|---|---|---|---|---|---|
| $n$ | orbital | $\epsilon_n$ | $z_n$ | $E^N_{\rm DFT}(n)$ | $\sum_{m=0}^{n} E^N_{\rm DFT}(m)$ | $\epsilon_n$ | $z_n$ | $E^N_{\rm DFT}(n)$ | $\sum_{m=0}^{n} E^N_{\rm DFT}(m)$ |
| 0 | n.a. | n.a. | 0.000 | $-1.5_{10}{+6}$ | $-1.5_{10}{+6}$ | n.a. | 0.000 | $-2.9_{10}{+6}$ | $-2.9_{10}{+6}$ |
| 1 | $1s_{1/2}$ | −35.181 | 0.021 | $1.5_{10}{+6}$ | 0.178 | −36.985 | 0.038 | $2.9_{10}{+6}$ | 731.008 |
| 2 | $1p_{3/2}$ | −20.302 | 0.045 | 0 | 0.178 | −20.691 | 0.077 | $-8.6_{10}{+2}$ | −125.448 |
| 3 | $1p_{1/2}$ | −15.040 | 0.072 | $-1.4_{10}{+2}$ | −140.540 | −14.783 | 0.131 | $-1.7_{10}{+2}$ | −142.584 |
| 4 | $1d_{5/2}$ | −6.528 | 1.429 | 0 | −140.540 | −6.399 | 1.462 | $1.2_{10}{-1}$ | −142.464 |
| 5 | $2s_{1/2}$ | −2.166 | 11.255 | $-1.1_{10}{+3}$ | −1235.194 | −2.685 | 5.702 | $-4.8_{10}{+1}$ | −190.628 |
| 6 | $1d_{3/2}$ | 2.831 | 17.458 | 0 | −1235.194 | 3.143 | 10.952 | $4.0_{10}{+1}$ | −150.627 |
| 7 | $1f_{7/2}$ | 10.265 | 31.398 | 0 | −1235.194 | 10.325 | 19.033 | $2.5_{10}{+1}$ | −125.608 |

The spurious contributions may result in large errors in the projected PNP energies, making results of the standard PNP calculations meaningless. Unfortunately, this is true not only for Skyrme forces that use density-dependent terms of fractional orders but also for the Gogny force, which contains a density-dependent term of order $\gamma = 1/3$.

### C. Calculations for deformed nuclei

In our previous study [44], we calculated the complete HFB mass chart of even-even nuclei by performing the PNP of paired ground states determined by the LN method. At this point, when performing the PNP calculation in each individual nucleus, one should take care of the cases when one of the poles $z_n$ turns out to be near the standard $r_0 = 1$ integration circle (unit circle).

In order to produce the ground state masses for all even-even nuclei lying between the two-nucleon drip lines, one has to calculate about 6000 nuclei. Moreover, each nucleus has to be calculated three times, by starting from oblate, spherical, and prolate initial shapes. We have found that among these 6000 nuclei, about 100 have a neutron or proton state with occupation numbers near 1/2. Therefore, the standard PNP method yields about 100 questionable results across the mass chart. However, the situations is much more serious when performing the constrained HFB calculations discussed in the following sections.

### D. Distribution of poles as a function of deformation

When increasing the quadruple deformation, states with smallest (largest) angular momentum projections onto the symmetry axis, $\Omega$, become more (less) bound on the prolate side, and the opposite holds for the oblate side. For states located above the shell gap, this means that low-$\Omega$ and high-$\Omega$ orbitals become more occupied with increasing prolate and oblate deformation, respectively. Therefore, at some deformation, these orbitals cross the Fermi energy and the corresponding poles cross the unit circle. An analogous situation may also occur for orbitals located below the shell gap, whereupon high-$\Omega$ and low-$\Omega$ Nilsson orbitals become less occupied with increasing prolate and oblate deformation, respectively, and also may cross the Fermi energy. We wish to emphasize that the problem occurs not at the point where the orbitals from above and below the shell gap cross each other, leading to a configuration change, but at deformation where either of these orbitals crosses the Fermi energy.

Such a case is illustrated in Fig. 7 for the nucleus $^{18}$O. In pure HFB calculations (no LN correlations included), this nucleus has neutron pairing only. At the spherical shape, the $1d_{5/2}$ shell is located above the $N$=8 shell gap, i.e., it has particle character ($|z| > 1$). The threefold degeneracy of this shell ($q$=3) makes the contribution from this pole to the projected energy vanish. At nonzero deformations, however, the degeneracy is lifted and three individual poles ($q$=1) appear in the complex plane. Moreover, near $\beta = 0.12$ and $\beta = -0.12$, poles corresponding to the $\Omega$=1/2 and $\Omega$=5/2 Nilsson levels cross the unit circle $|z| = 1$.

The situation is much worse for nuclei having more single-particle states with poles close to the unit circle. The neutron-rich nucleus $^{32}$Mg is such a complicated case illustrated in Fig. 8. This example is calculated in the HFB+LN approach, in which both neutron and proton pairing is nonzero. For completeness, canonical single-

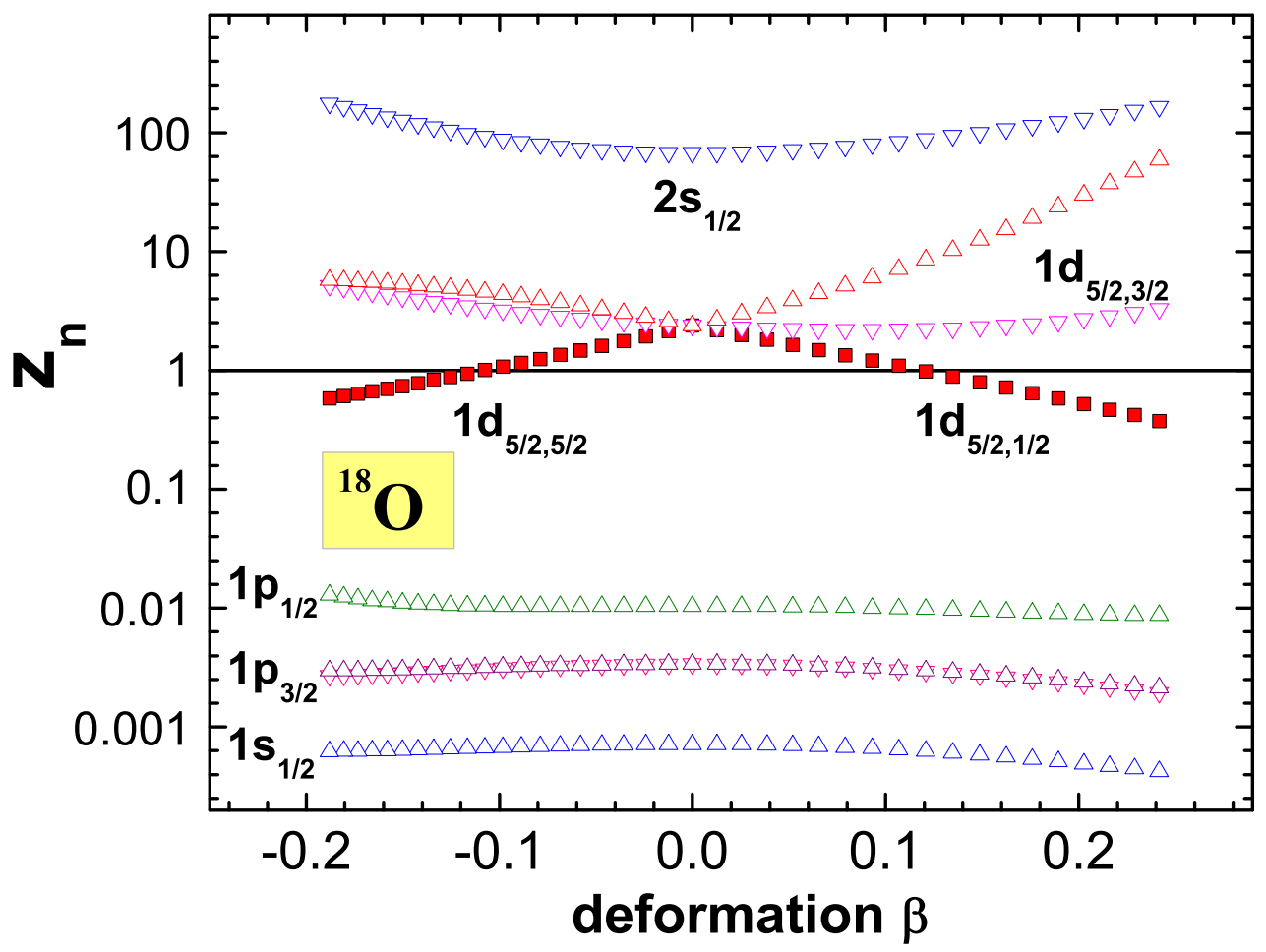

FIG. 7: (color online) Neutron poles $z_n$ in $^{18}$O (28) as functions of quadrupole deformation $\beta$, calculated within the HFB-SIII method with volume pairing interaction.

particle energies $e_n$ associated with the poles $z_n$ are plotted in Fig. 9

As can be seen in Fig. 8, there appear numerous crossings of poles with the unit circle as a function of deformation. On the prolate side, neutron poles $1f_{7/2,\Omega=1/2}$ (Nilsson level [330]1/2) and $1d_{3/2,3/2}$ ([202]3/2) cross the unit circle at the same deformation where they cross one another. At larger deformation, the same situation occurs for the $1f_{7/2,3/2}$ ([321]3/2) and $1d_{3/2,1/2}$ ([200]1/2) orbitals. For protons, a single $1d_{5/2,5/2}$ ([202]5/2) orbital crosses the unit circle at small deformations. On the oblate side, neutron orbitals $1f_{7/2,5/2}$ ([312]5/2) and $1d_{3/2,1/2}$ cross the unit circle at different deformations, near the point where they cross one another, while the proton $1d_{3/2,1/2}$ orbital stays near the unit circle for a wide range of deformations.

As discussed in Secs. III C and IV B, results of the PNP, at least for the density-independent terms, must only depend on the residues of poles that are inside the integration radius $r_0$. However, whenever a given pole crosses the integration contour, the projected energy must undergo a sudden jump as a function of deformation. This jump is, of course, equal to the residue at this pole. The fact that a given pole crosses the integration contour could be without consequence, provided the contour is shifted back to always stay between the same poles. This is always possible, as long as the poles do not cross around the contour. It is obvious that whenever they do, the projected energy may have a sudden jump that cannot be avoided by a contour shift. On the other hand, when two poles cross precisely at the integration contour, the corresponding degeneracy factor $q$ increases by a unity, and the poles may simply disappear (at least for the terms that show polynomial density dependence), in which case the projected energy may stay smooth. Such cases are studied in the next section.

### E. Deformation energy within the HFB+PNP method

Results in this section were obtained with the SIII Skyrme force whose density-dependent terms do not create additional problems (cf. Sec. IV). Let us first analyze the simpler case of $^{18}$O. Figure 10 presents the deformation energy $E(\beta)$ as a function of the quadrupole deformation $\beta$. As a reference curve, we show the unprojected deformation energy emerging from the HFB calculations and the associated PNP energy curve (PAV; solid squares; the contour radius $r_0$=1). Near the ground state ($\beta = 0$), the projected energy is lowered by about 1.5 MeV due to additional PN correlations. At larger deformations, the correlation energy decreases due to the stronger static neutron pairing.

At deformations $\beta \simeq -0.12$ and $\beta \simeq +0.12$, the projected energy curve exhibits unphysical jumps. By comparing with Fig. 7, one concludes that at these deformations the neutron $1d_{5/2}$ poles cross the integration contour. Obviously, the residue contributions of these poles cause the sudden jumps in the deformation energy. The $1d_{5/2,5/2}$ pole introduces a positive contribution at $\beta \simeq -0.12$, while the $1d_{5/2,1/2}$ pole introduces another positive contribution at $\beta \simeq +0.12$. Based on this observation, two other sets of PAV calculations were carried out. The first calculation (open circles) was done by excluding contributions from the $1d_{5/2}$ poles, as is the case for the ground state configuration. This can be accomplished by reducing the integration radius from $r_0 = 1$ to a smaller value of about $r_0 = 0.1$, cf. Fig. 7. At small deformations, $-0.12 \leq \beta \leq 0.12$, the new results are identical to those obtained with the unit circle, and at the larger deformations (prolate or oblate), the energy curve smoothly continues without any jump. Thus, in this example, an appropriate shift of the integration contour allows us to obtain smooth and unique projected energy. The second PAV curve (open squares) has been obtained by always including the lowest $1d_{5/2}$ pole, i.e., by continuously varying the integration radius as a function of $\beta$, to ensure that it always stays between the first and second $1d_{5/2}$ pole (cf. Fig. 7). The resulting energy curve coincides at large deformations with the standard PAV result, and then smoothly continues to $\beta = 0$, where the $1d_{5/2}$ poles ($q$=3) disappear.

Figure 10 also presents the fully self-consistent VAP results. Similar to the PAV case, two sets of calculations were performed. The solid (open) stars correspond to including (excluding) the contributions from the lowest $1d_{5/2}$ poles. In both cases, one obtains smooth curves, which, beyond the spherical point, differ from one another.

In this rather simple case of $^{18}$O, both in the PAV and VAP calculations, one can avoid unphysical jumps of the projected energy curve by making a specific selection of "active" poles that are considered during contour integration. Such selection of residues can, in principle, become a part of the definition of the projected energy.

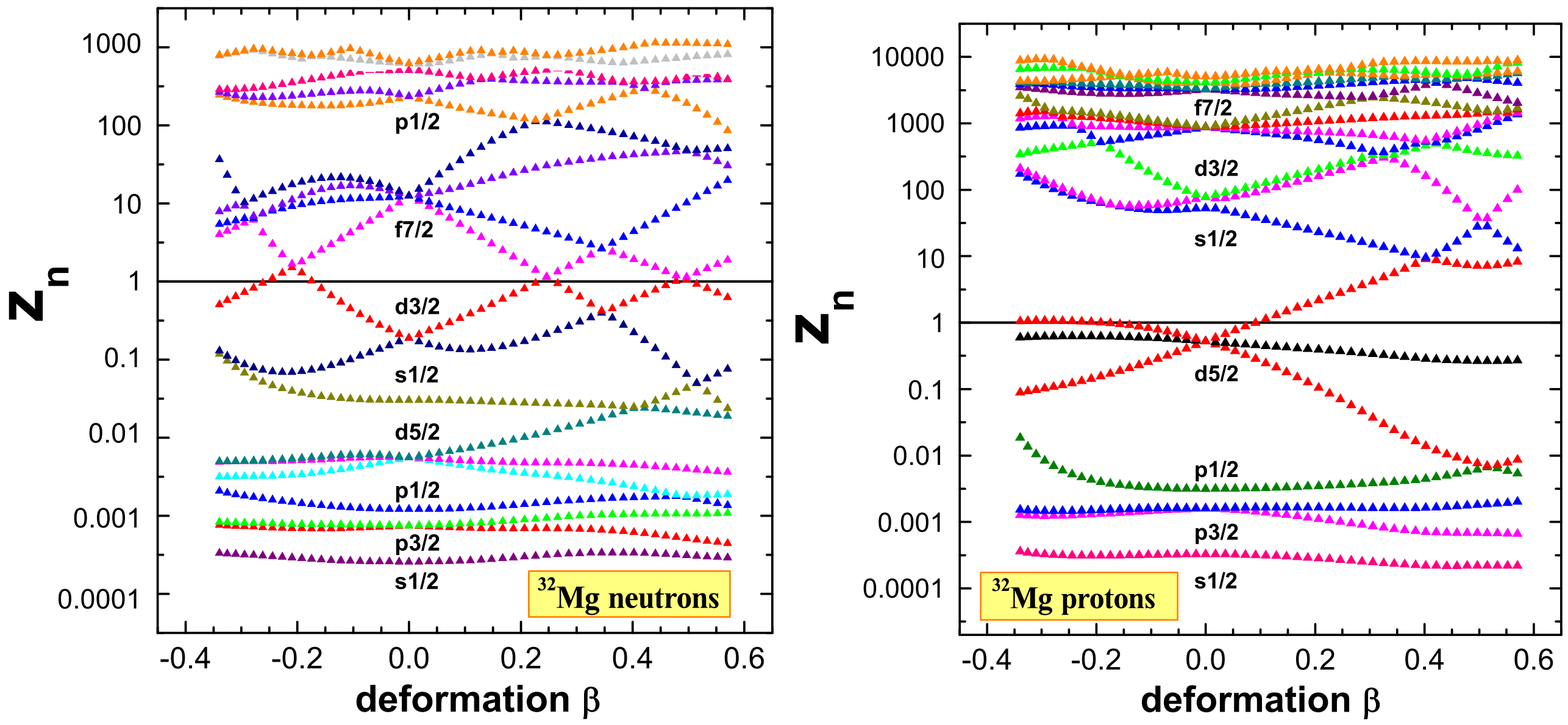

FIG. 8: (color online) Neutron (left) and proton (right) poles $z_n$ (28) as functions of quadrupole deformation $\beta$ calculated for $^{32}$Mg with the SIII functional and volume pairing interaction.

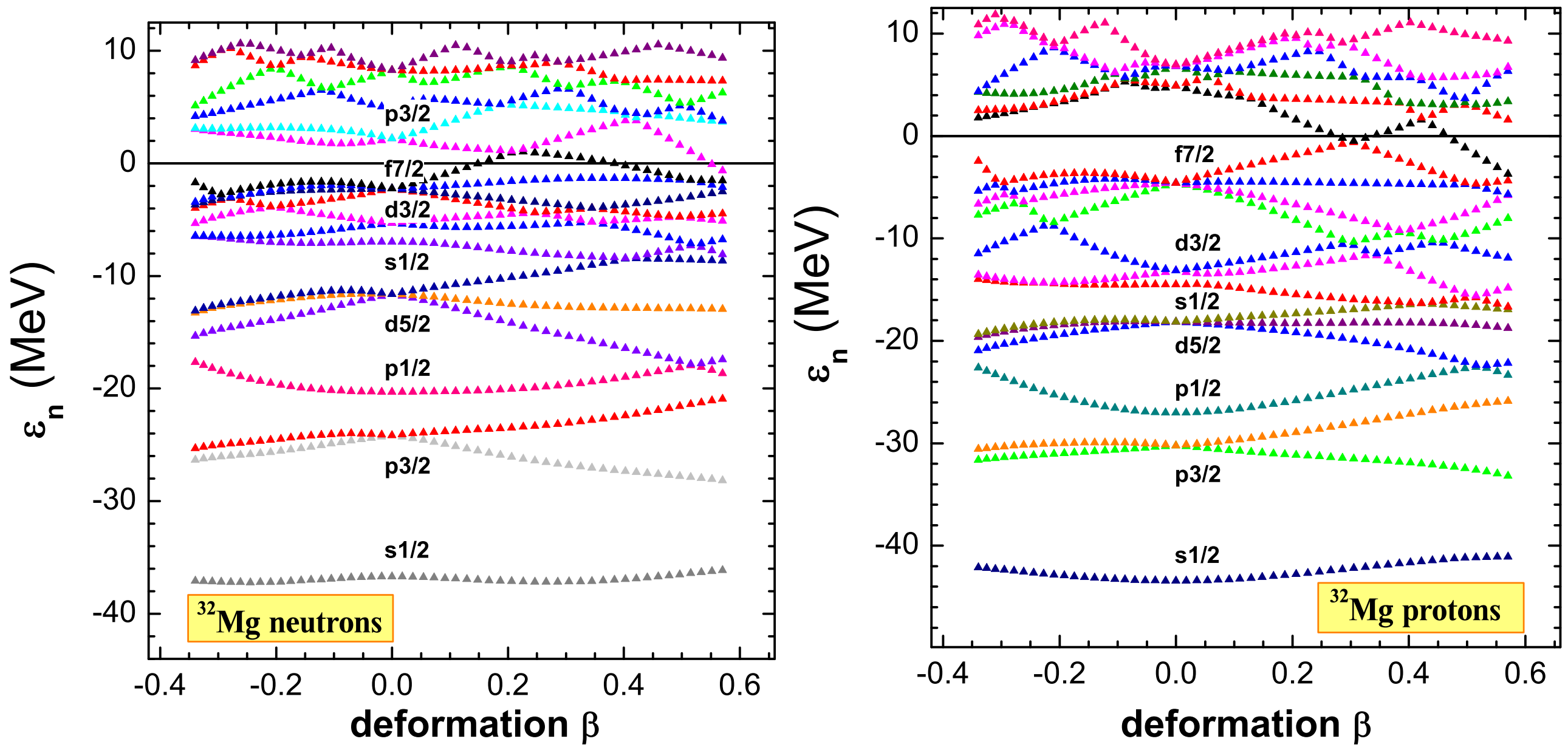

FIG. 9: (color online) Similar to Fig. 8 except for canonical energies $e_k$.

The variational principle can then be invoked to pick the selection that yields the lowest projected energy. In the discussed case of $^{18}$O, the PAV and VAP energies obtained by excluding the 1d$_{5/2}$ poles are the lowest, and they are smooth functions of deformation. Therefore, such a selection can be adopted for the final PNP energy in this nucleus. It is clear, however, that one cannot *a priori* tell which selection of poles leads to the lowest projected energy. For example, in heavier oxygen isotopes, the lowest energy is obtained by including some of the 1d$_{5/2}$ poles.

Let us now consider a more complicated case of the HFB+LN calculations for the neutron rich nucleus $^{32}$Mg. The total HFB energy (without the corrective $\lambda^{(2)}$ LN term) is shown in Fig. 11 as a function of $\beta$. Solid squares denote the result of PAV PNP calculations on the top of HFB+LN. At $\beta \simeq +0.1$, the PAV curve exhibits a small jump, after which its behavior changes character. This is clearly related to the proton 1d$_{5/2,5/2}$ pole crossing the unit circle, cf. Fig. 8. Otherwise, the PAV deformation energy is quite smooth as a function of deformation, despite the fact that three pairs of neutron poles cross the unit circle in the deformation range considered. This apparent lack of sensitivity to neutron poles can be traced back to the fact that they cross the integration contour at or near points where they pairwise cross one another. Therefore, the increasing degeneracy factor $q$ makes the poles disappear at the crossing points; hence, the total PAV curve behaves smoothly.

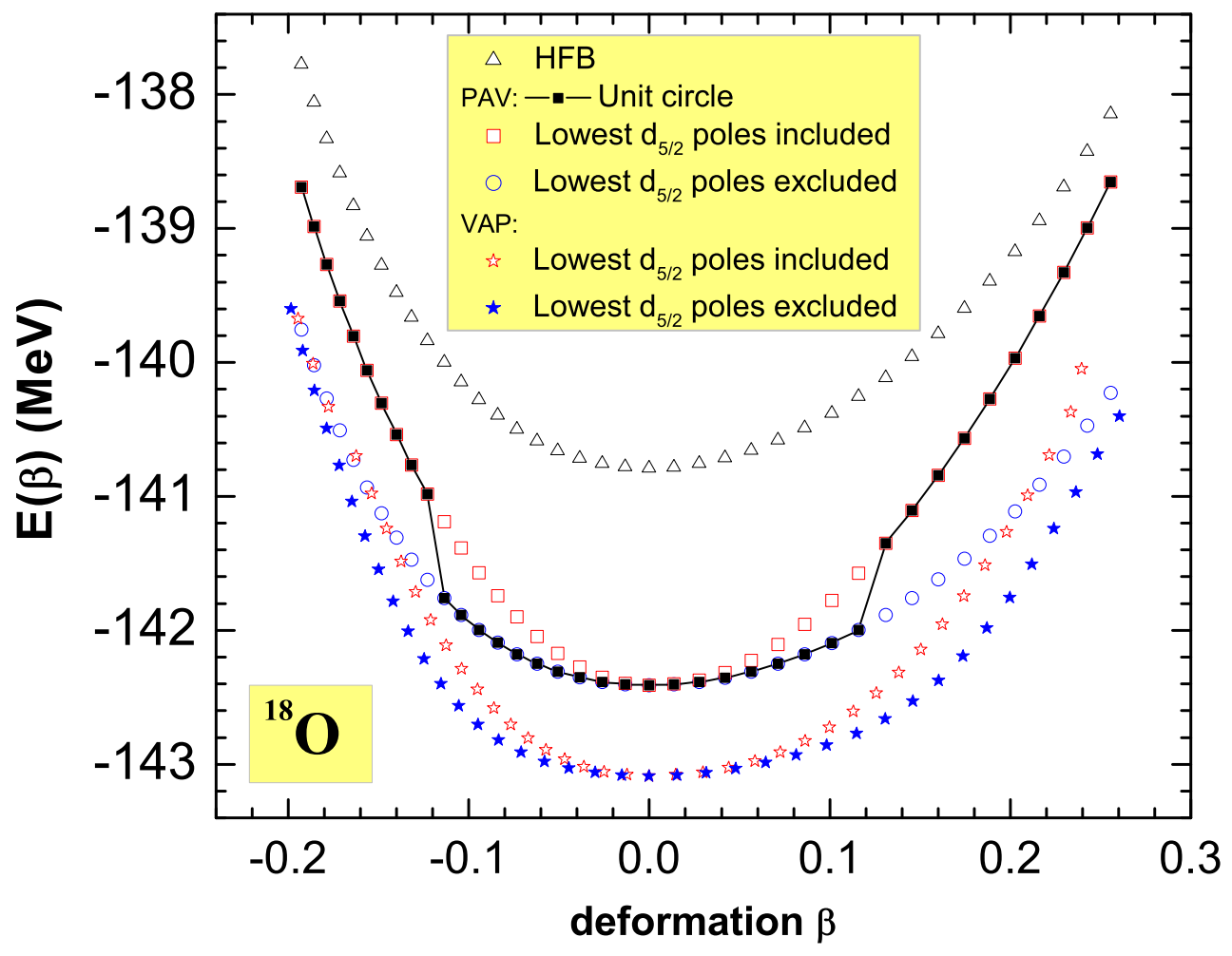

FIG. 10: (color online) Deformation energy $E(\beta)$ as a function of quadrupole deformation $\beta$, calculated for $^{18}$O with the SIII force and volume pairing interaction. Results of HFB (open triangles) are compared with different variants of PAV (squares and circles) and VAP PNP (stars) (see text for details).

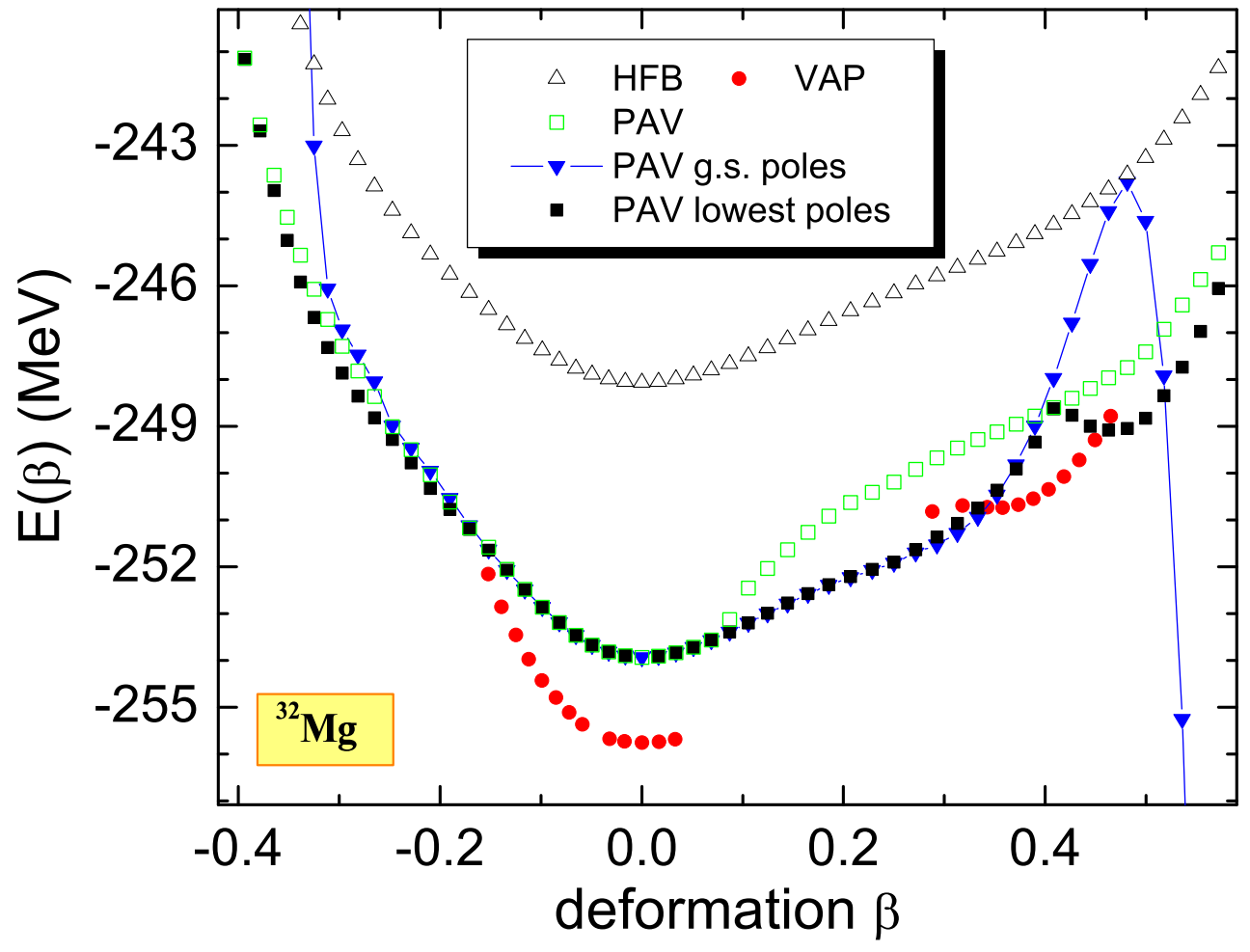

FIG. 11: (color online) Deformation energy $E(\beta)$ as a function of quadrupole deformation $\beta$ calculated for $^{32}$Mg with the SIII force and volume pairing interaction. Results of the PAV HFB+LN calculations (squares and triangles) are compared with the VAP PNP results (dots). The standard HFB result is shown by open triangles.

Following the example of $^{18}$O, also for $^{32}$Mg we performed PAV calculations wherein we took into account contributions from all the poles that contribute to the ground state configuration ($\beta = 0$). The resulting PAV curve (solid triangles in Fig. 11) behaves smoothly but shows unphysical behavior at very large deformations. An explanation of this artifact follows from Fig. 12, where we plot energy contributions from the most important poles: (i) the neutron $1d_{3/2,3/2}$ pole (solid circles; it leaves the unit-circle at $\beta \simeq 0.22$ and we have to add its contribution beyond this point); (ii) the neutron $1d_{7/2,1/2}$ pole (solid squares; it enters the contour at $\beta \simeq 0.22$ and we have to subtract its contribution beyond this point); and (iii) the proton $1d_{5/2,5/2}$ pole (solid triangles; it leaves the contour at $\beta \simeq 0.1$ and we have to add its contribution beyond this point). Interestingly, pole contributions to the total projected energy oscillate with deformation. As expected, the residues vanish when the corresponding poles cross at the integration circle, cf. Fig. 8. However, oscillations between the crossing points can become quite large, as is the case for the proton $1d_{5/2,5/2}$ pole; hence, the projected energy curve obtained by keeping contributions from the ground-state set of poles acquires strong unphysical oscillations.

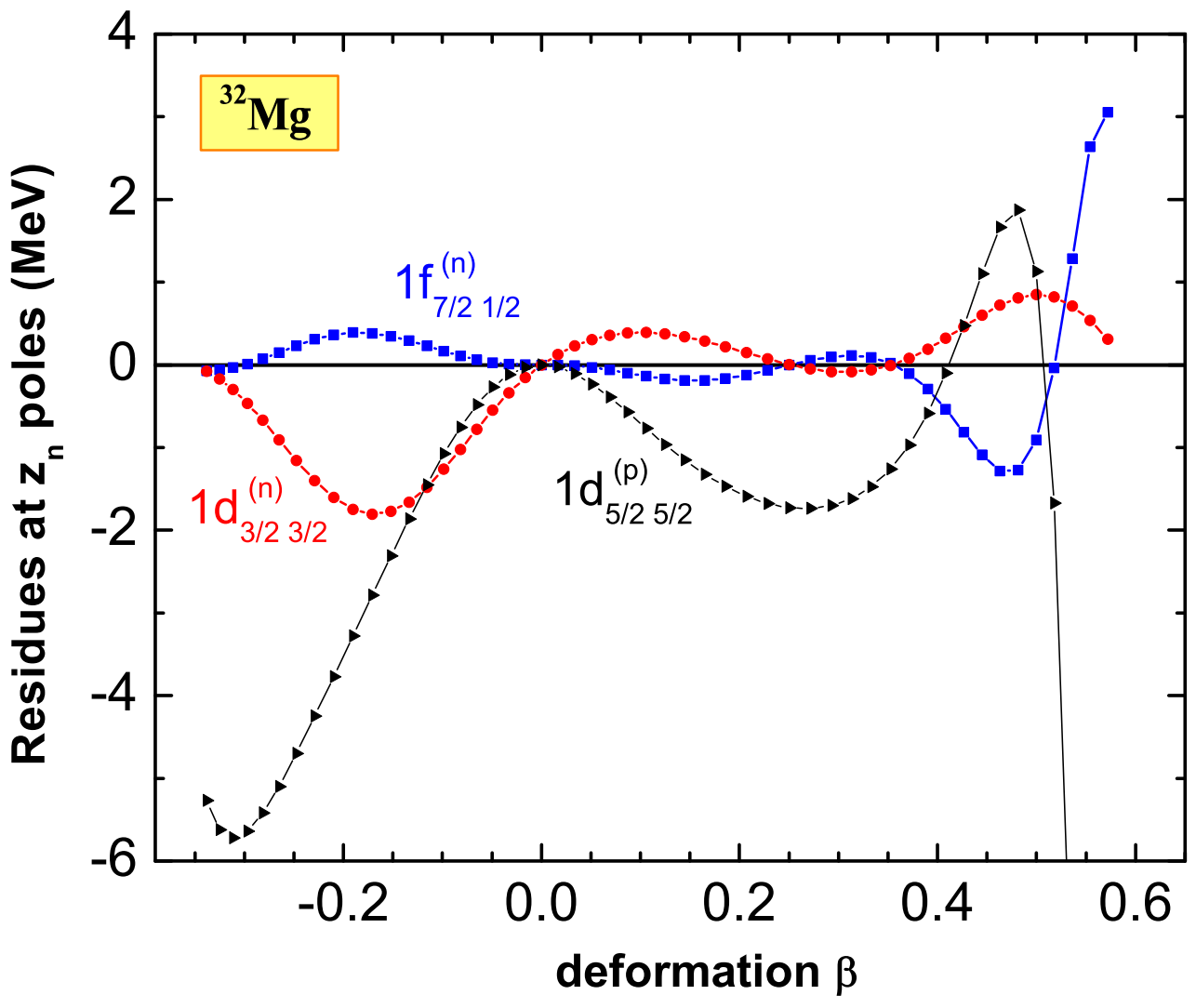

FIG. 12: (color online) Contributions to the PAV energy of $^{32}$Mg from the three selected poles as a function of deformation $\beta$.

In a search for the most sensible method of calculating the projected energy curve within the PAV approach, we can employ a prescription whereby the number of the lowest poles is kept fixed within the integration contour. This can be realized by keeping $r_0$ between the poles or at the pole crossing point. Since at the crossing points the poles vanish, at least in SIII calculations, this results in a smooth energy curve. Such an option is shown in Fig. 11 with solid squares. The resulting curve is indeed very smooth; however, at $\beta \simeq 0.4$ there appears an unphysical bump, which makes this option as unacceptable as the other one.

We have also attempted calculating the energy curve within the VAP approach. In principle, the VAP approach could have generated problems related to the fact that the density-dependent terms of fractional order may lead to large negative contributions (see Sec. III F). In practice, this is never the case because the VAP method [23] is not implemented through an explicit minimization of the energy, but is carried out by solving variational equations that have been derived with the same incorrect treatment of cuts in the complex plane. In this context, it is worth emphasizing that the appearance of poles never leads to infinite total energies, but to discontinuities in the total energy. Therefore, there is no danger that the minimization procedure may attract a solution towards a pole.

The main problem in implementing the VAP method is related to the fact that unprojected quantities, e.g., particle $\rho_{nn'}$ or pairing $\tilde{\rho}_{nn'}$ densities, lose their usual physical meaning [23] in VAP. They depend on the internal normalization $N' = \mathrm{T}r\rho$ of the density $\rho_{nn'}$ that is not related to the particle number $N$ onto which the state is projected. Neither the total VAP energy nor other projected observables depend on the normalization $N'$. However, depending on the choice of the internal normalization $N'$, one obtains different canonical occupation probabilities; hence, the associated poles $z_n$ are not distributed in the same way as in the unprojected HFB case. Depending on the internal normalization $N'$, different poles $z_n$ enter the integration contour, and the convergence procedure cannot be easily controlled.

Additional problems arise when two poles are nearly degenerate. Although at the point of degeneracy the poles disappear, when the distance between the poles is small but nonzero and the integration contour is between them, one faces significant instabilities of the constrained VAP problem. During the iteration of VAP equations, one or both poles enter or leave the integration contour. The poles create jumps in the projected energy and, which is even more important, they create jumps in the deformation. The numerical algorithm enters a 'ping-pong' regime, which cannot be overcome, and one cannot converge to any solution. Figure 11 offers a good illustration of this problem. The converged VAP energies for $^{32}$Mg are shown with solid circles. The converged solution can be found only in limited regions of deformation. The 1d and 1f neutron poles close to the contour spoil the convergence in the regions of $\beta \approx -0.2$, 0.25, and 0.5. The same is true for the proton 1d states around $\beta \approx -0.2$ and 0.1. As a result, the VAP procedure could be solved only within small deformation intervals around $\beta \approx -0.2$, 0, and 0.4.

## V. CONCLUSIONS

This study contains the systematic analysis of the particle-number-projected DFT approach. This approach, usually in the Skyrme-HFB or Gogny-HFB framework, is commonly used in systematic calculations of nuclear ground-state properties, low-energy excitations, and high-spin states. For heavy, complex nuclei, the nuclear DFT is the only viable microscopic tool based on the effective interaction (or functional). To advance the nuclear DFT further, to improve its theoretical foundations, and to make it a more reliable tool, it is important to fully understand the advantages and drawbacks of the method when applied to self-bound nuclear systems.

The main conclusions of this work can be summarized as follows:

1. The transition density matrices connecting states having different orientation in the gauge plane $z$ have poles on the imaginary axis Im[$z$]. In the HFB formalism that is based on one Hamiltonian acting in all channels (Hamiltonian-based HFB), these poles are irrelevant as their impact is nullified by the cancellation between the Hamiltonian matrix elements originating from particle-hole and pairing channels. Such a cancellation is not present in the HFB applications in which some of the Hamiltonian matrix elements are neglected (or approximated), in the HFB method based on density-dependent interactions (usually acting in different channels), and in the DFT approach in which the Hamiltonian does not appear at all. In all those cases, the projection operator is not defined uniquely and the result depends on the analytic structure of the transition energy density; hence, the projected DFT energy depends on the integration contour. The resulting PNP energy can be expressed in terms of individual residues corresponding to the poles $z_n$ associated with single-particle (canonical) proton and neutron orbits. In contrast, in the Hamiltonian-based HFB, the result depends only on the single pole at the origin ($z_0 = 0$).

2. Within the Hamiltonian-based HFB, there exist sum rules that relate the unprojected matrix elements with the matrix elements in the projected states. Similar sum rules can be derived within the DFT; they relate the unprojected DFT transition energy density with the projected DFT energies. The DFT sum rules offer the interpretation of the projected DFT energies as Fourier components (associated with irreducible representations of gauge group U(1)) of the DFT transition energy density. This can be naturally extended to other (higher) broken symmetry groups, such as SU(2) (associated with the broken angular momentum symmetry).

3. The discussion of the particle number restoration can be extended to other symmetry restoration problems. In particular, DFT transition densities associated with angular-momentum-projected states are expected to have complicated pole structure in the three dimensional space of Euler angles (see the example shown in Ref. [45]).

4. For the terms in the density functional that have

polynomial density dependence, the appearance of poles inside the contour gives rise to sudden jumps of the projected energy whenever the contour's pole content changes. Otherwise, the results are stable. This is not true for the terms having fractional-power density dependence (e.g., density-dependent pieces of many Skyrme and Gogny interactions or the Coulomb exchange term taken in the Slater approximation). Here, the dependence on the contour radius shows a strong subthreshold behavior that can only be cured by considering appropriate integration contours which do not go across the cuts in the complex $z$-plane. Other prescriptions give rise to uncontrolled energy behavior resulting from the fact that the corresponding integration contours do not close.

5. As a practical measure that allows avoiding problems related to the fractional-power density dependence, we propose using the integration contours which pass near and above the poles. Although such a prescription requires using rather dense meshes of integration points, it minimizes the risks of crossing the cuts in the complex plane. In this way, the ambiguities related to the non-analyticity of the DFT transition energy are reduced to those corresponding to the choice of poles included within the integration contour.

6. Projected DFT yields questionable results if a pole appears very close to the integration contour. While such a situation seldom happens in the ground-state calculations (less than 2% cases are affected), it frequently occurs in calculations of projected energy surfaces, such as those in the generator coordinate method (GCM). The appearance of poles in the vicinity of the contour as a function of the collective coordinate (e.g., deformation) gives rise to uncontrolled irregularities and jumps in the results; in particular, it makes it impossible to define the PNP potential energy surfaces.

7. Pole pathologies appear in a particularly strong way in the fully self-consistent VAP calculations. In this approach, transition density poles are not uniquely defined; moreover, their positions can change during the iteration process leading to numerical instabilities.

8. The analytic structure of the transition energy density becomes exceedingly complicated in nuclei with protons and neutrons paired, thus requiring simultaneous proton and neutron PNP. Of particular importance in the context of GCM applications is the extension of the present analysis to non-diagonal matrix elements between the PNP states.

Some of the problems listed above, in particular those related to the configuration-mixing DFT method and applications of the generalized Wick's theorem to DFT, have been recently addressed in a series of papers [33, 34]. In these studies, a practical cure has been proposed that is based on removing specific spurious components of the DFT functional that can be associated with self-interaction and self-pairing. When applied to truncated Hamiltonians, this practical prescription turns out to be very effective [46]. However, this kind of solution does not remove ambiguities related to using complicated (e.g., fractional-power) dependence of the energy density functionals on particle densities. Finding ultimate cures to the problems discussed in our study will undoubtedly result in establishing better theoretical constraints on the form of the DFT energy density functionals for nuclear self-bound systems.

## Acknowledgments

This work was supported in part by the Polish Ministry of Science; by the Academy of Finland and University of Jyväskylä within the FIDIPRO programme; by the U.S. Department of Energy under Contract Nos. DE-FG02-96ER40963 (University of Tennessee), DE-AC05-00OR22725 with UT-Battelle, LLC (Oak Ridge National Laboratory), DE-FG05-87ER40361 (Joint Institute for Heavy Ion Research), and DE-FC02-07ER41457 (University of Washington); and by the National Nuclear Security Administration under the Stewardship Science Academic Alliances program through DOE Research Grant DE-FG03-03NA00083.